\documentclass[usegraphicx,usenatbib,useapjfonts,apj]{emulateapj}
\usepackage{amssymb}
\usepackage{aas_macros}

\usepackage{graphicx} 

\usepackage{psfrag,color}
\usepackage{psfig}
\usepackage{times}

\usepackage{longtable}

\def\lcdm{{$\Lambda$CDM}}
\def\mpch{\mbox{$h^{-1}$Mpc}}

\def\msun{\mbox{$M_\odot$}}

\def\mh{\mbox{$M_{v}$}}
\def\rh{\mbox{$R_{\rm v}$}}

\def\ms{\mbox{$M_{s}$}}
\def\mg{\mbox{$M_{g}$}}
\def\mb{\mbox{$M_{b}$}}
\def\mbv{\mbox{$M_{b}^{\rm v}$}}
\def\msv{\mbox{$M_{s}^{\rm v}$}}
\def\fg{\mbox{$f_{g}$}}

\def\Fb{\mbox{$F_{b}$}}
\def\Fs{\mbox{$F_{s}$}}
\def\Fsv{\mbox{$F_{s}^{\rm v}$}}
\def\Fbv{\mbox{$F_{b}^{\rm v}$}}

\def\FbU{\mbox{$F_{b,U}$}}
\def\ome{\mbox{$\Omega_m$}}
\def\omem{\mbox{$\Omega_m$}}
\def\omel{\mbox{$\Omega_\Lambda$}}
\def\omeb{\mbox{$\Omega_b$}}

\def\re{\mbox{$R_{e}$}}

\def\vmax{\mbox{$V_{\rm max}$}}
\def\vcir{\mbox{$V_{\rm C}(R)$}}

\def\ssfr{\mbox{sSFR}}
\def\zh{\mbox{$z_{f,h}$}}
\def\zg{\mbox{$z_{f,s}$}}

\def\mathnew{\mathsurround=0pt}
\def\simov#1#2{\lower .5pt\vbox{\baselineskip0pt
    \lineskip-.5pt\ialign{$\mathnew#1\hfil##\hfil$\crcr#2\crcr\sim\crcr}}}

\def\simless{\mathrel{\mathpalette\simov <}}

\def\cm3{\mbox{cm$^{-3}$}}
\def\Ke{\mbox{K}}
\newcommand{\nsf}{n_{\rm SF}}
\newcommand{\rhosf}{\rho_{\rm SF}}
\newcommand{\Tsf}{T_{\rm SF}}

\def\apjl{ApJ}%
\def\apjs{ApJS}%
\def\ao{Appl.~Opt.}%
\def\aap{A\&A}%
\def\spose#1{\hbox to 0pt{#1\hss}}
\newcommand\lsim{\mathrel{\spose{\lower 3pt\hbox{$\mathchar"218$}}
     \raise 2.0pt\hbox{$\mathchar"13C$}}}
\newcommand\gsim{\mathrel{\spose{\lower 3pt\hbox{$\mathchar"218$}}
     \raise 2.0pt\hbox{$\mathchar"13E$}}}
 
\newcommand{\bc}{\begin{center}}
\newcommand{\ec}{\end{center}}

\slugcomment{Submitted}

\shorttitle{Simulations of isolated dwarf galaxies}
\shortauthors{Gonz\'alez-Samaniego et al.}

\begin{document}

\title{Simulations of isolated dwarf galaxies formed in dark matter halos with different mass assembly histories}
\author{A. Gonz\'alez-Samaniego\altaffilmark{1}, P. Col\'{\i}n\altaffilmark{2}, V. Avila-Reese\altaffilmark{1},
A. Rodr\'{\i}guez-Puebla\altaffilmark{1}, and O. Valenzuela\altaffilmark{1}
}

\altaffiltext{1}{Instituto de Astronom\'{\i}a, Universidad Nacional Aut\'onoma de M\'exico, 
A.P. 70-264, 04510, M\'exico, D.F., M\'exico}

\altaffiltext{2}{Centro de Radioastronom\'{\i}a y Astrof\'{\i}sica, Universidad Nacional 
Aut\'onoma de M\'exico, A.P. 72-3 (Xangari), Morelia, Michoac\'an 58089, M\'exico }


\begin{abstract}
We present zoom-in {\it N}-body/hydrodynamics resimulations of dwarf galaxies formed in isolated cold dark matter (CDM) halos with {\it the same virial mass} ($\mh\approx2.5\times 10^{10}\msun$) at redshift $z=0$. Our goals are to (1) study the mass assembly histories (MAHs) of the halo, stellar, and gaseous components; and (2) explore the effects of the halo MAHs on the stellar/baryonic assembly of simulated dwarfs.
Overall, the dwarfs are roughly consistent with observations. More specific results include: (1) the stellar-to-halo mass ratio remains roughly constant since $z\sim 1$, i.e., the stellar MAHs closely follow halo MAHs.
(2) The evolution of the galaxy gas fractions, \fg, are episodic, showing that the supernova-driven outflows play an important role in regulating \fg\ and hence, the star formation rate, SFR; however, in most cases, a large fraction of the gas is ejected from the halo.
(3) The star formation histories are episodic with changes in the SFRs, measured every 100 Myr, of factors 2--10 on average.
(4) Although the dwarfs formed in late assembled halos show more extended SF histories, their $z=0$ specific SFRs are still below observations. (5) The inclusion of baryons most of time reduces the virial mass by 10\%--20\% with respect to pure $N$-body simulations. Our results suggest that rather than increasing the strength of the supernova-driven outflows, processes that reduce the star formation efficiency could help to solve the potential issues faced by CDM-based simulations of dwarfs, such as low values of the specific SFR and high stellar masses.
\end{abstract}
                                                                                                                 
\keywords{dark matter -- galaxies: dwarf -- galaxies:evolution -- galaxies: formation -- galaxies: halos -- methods: numerical}


\section{Introduction}
\label{intro}

The $\Lambda$ cold dark matter (\lcdm) cosmology provides a robust theoretical background for understanding
galaxy formation and evolution.  The properties and evolution of galaxies predicted by the results of 
models and simulations based on the \lcdm\ cosmology are encouraging, though several potential issues 
remain yet to be solved. One of these potential issues refers to the history and efficiency of the stellar mass assembly 
of low-mass galaxies, those with stellar masses $\ms\lesssim 10^{10}$ \msun. 
 
Several observational pieces of evidence show that low and intermediate redshift galaxies less massive than 
$\ms\sim (5-10) \times 10^{9}$ \msun\ have high specific star formation rates ($\ssfr\equiv$ SFR/\ms), 
with larger values, on average, for less massive galaxies \citep[e.g.,][]{Salim+2007,Noeske+2007b,Rodighiero+2010,Karim+2011,
Bauer+2011,Gilbank+2011,Bauer+2013}. 
This shows that the smaller the galaxies, the later they assemble their stellar mass, on average, having younger stellar
populations (downsizing in \ssfr; see for reviews \citealp{Fontanot+2009}; \citealp{Firmani+2010b}).  From an analysis of local 
galaxies, \citet{Geha+2012} conclude that virtually all galaxies less massive than $\ms\approx 10^9$ \msun\ 
in the field show evidence of high recent star formation (SF) and have relatively young stellar populations.     
On the other hand, by applying a Bayesian analysis of the observed spectral 
energy distribution (SED) of low-mass galaxies ($\ms=1.6-4.0\times 10^9$ \msun) at $0.2<z<1.4$ with synthetic 
SEDs, \citet{Pacifici+2013} find that these galaxies have, on average, a rising SF history (SFH), contrary to massive 
galaxies for which the SF decreases with time \citep[see also][]{Perez+2013}. 
Finally, it should be said that low-mass galaxies, specially dwarf galaxies, are susceptible to 
episodic (bursty) SFH; hence, any statistical inference of an average SFH could be 
biased \citep{Bauer+2013}.

In agreement with the apparently late \ms\ assembly, several 
pieces of evidence show that low-mass galaxies have very low stellar and baryonic mass fractions, 
$\Fs \equiv \ms/\mh$ and $\Fb \equiv \mb/\mh$, respectively, 
where \mh\ is the virial halo mass: \Fs\ and \Fb\ exhibit a strong 
dependency on \mh, decreasing as \mh\ gets smaller.
\citep[e.g.,][]{Conroy+2009b,Guo+2010,Behroozi+2010,Behroozi+2013,Moster+2010,
Moster+2013,Rodriguez-Puebla+2011,Rodriguez-Puebla+2013,Papastergis+2012}.

By applying the common recipes and schemes for SF and feedback, most semi-analytic models 
\citep[e.g.,][]{Somerville+2008,Fontanot+2009,Santini+2009,Liu+2010,Bouchet+2010,Weinmann+2012} and high-resolution hydrodynamic
simulations \citep[e.g.,][]{Colin+2010,Sawala+2011,Avila-Reese+2011b,deRossi+2013} predict, instead, that
present-day low-mass galaxies, both satellites and centrals, are too red, passive, old, and efficient 
forming stars in the past, as compared to observations. In some recent works, it is discussed that these 
discrepancies can be reduced if a proper comparison of simulations with observations is used
\citep{Brook+2012, Munshi+2013} and/or when the H$_2$ formation process and a H$_2$-based
SF scheme are taken into account \citep{Kuhlen+2012,Christensen+2012,Munshi+2013,Thompson+2014}. 

As recently discussed in the literature, a key question for low-mass galaxies formed in the context of the 
\lcdm\ cosmology is whether the assembly of their stellar and halo masses are closely related or 
the former is systematically detached from the latter  
\citep[e.g.,][]{Conroy+2009a,Firmani+2010b,Leitner2012,Behroozi+2013,Yang+2012,Moster+2013,deRossi+2013}. 
These works suggest that the stellar mass assembly inferred from the average observed galaxy population, 
as compared to the average theoretical halo mass assembly history (MAH), shows an opposite trend in the sense 
that while the assembly of less massive \lcdm\ halos occurs earlier than the assembly of more massive ones,
the stellar mass assembly occurs later as the galaxy gets
smaller \citep[see, e.g., Figure 4 in][]{Firmani+2010b}.  
The complex baryonic physics of the galaxy evolution inside growing halos is summarized
in this behavior.

Thus, some important questions worth exploring in detail in high-resolution simulations of central low-mass galaxies are: 
What are the MAHs of the halo, stars, and gas in these galaxies? How much do the stellar/baryonic 
mass assembly and the galaxy properties depend on the different (stochastic) halo MAHs? 
How episodic (bursty) could their SFHs be? How much do the physics of baryons affect the dark matter (DM)
masses of the small halos at different epochs? These questions are addressed here using zoom
$N$-body/hydrodynamics simulations of seven distinct low-mass halos and their corresponding central galaxies, 
all with similar present-day halo masses (2-3)$ \times 10^{10} \msun$. The distinct halos are selected with the criterion of being relatively isolated in such a way that the central galaxies formed inside them can be associated 
with field dwarf galaxies. The study of subhalos/satellite galaxies have extra complications 
that are beyond the scope of the present work. 

In Section 2, we describe the code and main parameters that will characterize the simulations. 
In Section 3 we present several results from the simulations as a function of the halo MAHs. 
The role that baryons play in the halo mass assembly is studied in Section 4.
Section 5 is devoted to a discussion of the SF-driven outflows versus other processes that could delay
the SF in low-mass galaxies (Section 5.1), the analysis of where the baryons are in the simulated galaxies
(Section 5.2), and the episodic SFHs observed in the simulations (Section 5.3).   
Our conclusions are given in Section 6.   

\section{The method}
\label{sim}

\begin{table*}
 \begin{center}
  \caption{Physical properties at z = 0}
  \begin{tabular}{@{}cccccccccccc@{}}\\
  \hline
   Name   & log(\mh) &  log(\ms)\tablenote{Mass within 0.1\rh (the same applies for \mg).}  & log(\mg)   & \vmax  & \re\tablenote{Radius that encloses half of the stellar mass within 0.1\rh.}  & \rh  & \fg\tablenote{$\fg\equiv \mg/(\mg+\ms)$.}  &  $M_{g, \rm cold}/\mg$ & D / T\tablenote{Ratio of the mass contained in the high-angular momentum disk stars with respect to the total stellar mass.}  &\zh\tablenote{Redshift at which the given halo acquired one third of its present-day mass.}  & SFR  \\
         & (\msun) &(\msun)  &(\msun)   &(km $s^{-1}$)  &(kpc)  &(kpc) &  &  &  &  &($10^{-3} \msun$ $\rm yr^{-1}$) \\
  \hline
   Dw1    & 10.30  & 8.10   & 8.16  & 52.43  & 1.29  & 78.00 & 0.53 & 0.93 & 0.26 & 3.0  & 1.69  \\ 
   Dw2    & 10.46  & 8.35   & 8.27  & 49.59  & 1.57  & 79.16 & 0.45 & 0.47 & 0.33 & 2.3  & 0.87  \\ 
   Dw3    & 10.46  & 8.73   & 8.07  & 56.07  & 1.46  & 78.50 & 0.18 & 0.81 & 0.01  & 2.3  & 0.34  \\ 
   Dw4    & 10.38  & 8.38   & 8.61  & 52.87  & 1.14  & 73.64 & 0.63 & 0.67 & 0.19  & 1.9  & 1.08  \\  
   Dw5    & 10.46  & 8.55   & 9.08  & 61.52  & 1.57  & 77.34 & 0.77 & 0.82 & 0.49  & 1.9  & 1.69  \\ 
   Dw6    & 10.46  & 8.44   & 8.93  & 50.63  & 4.16  & 78.00 & 0.76 & 0.98 & 0.66  & 1.9  & 13.7  \\ 
   Dw7    & 10.39  & 8.21   & 8.60  & 43.44  & 3.14  & 73.90 & 0.71 & 0.75 & 0.59  & 1.7  & 7.76  \\  
  \hline
 \end{tabular}
 \label{table}
\end{center}
\end{table*}

\subsection{Code, Star Formation, and Feedback}
\label{the_code}

We perform numerical simulations with the adaptive mesh refinement  (AMR)
$N$-body/hydrodynamic ART code \citep[]{Kravtsov+1997,Kravtsov2003}. 
This is one of the few cosmological codes that uses the Eulerian method to solve 
the hydrodynamical equations for gas trapped in the DM cosmic structures. 
Other widely used AMR codes are ENZO
\citep{Bryan+1997} and RAMSES \citep{Teyssier+2002}. The ART code incorporates
a wide variety of physical processes, including:  metal, atomic and molecular 
cooling, homogeneous UV heating, metal advection, SF, and thermal feedback. 
The cooling and heating rates, which take into account Compton heating/cooling, 
and the UV heating from a cosmological background radiation \citep{Haardt+1996} are
tabulated for a temperature range of $10^2\ \Ke < T < 10^9\ \Ke$, and a grid of densities,
metallicities (from $Z=-3.0$ to $Z=1.0$, in solar units), and redshifts using the 
CLOUDY code \citep[ version 96b4]{Ferland+1998}. We set the minimum 
temperature in the code to 300 K; because of the absence of gas self-shielding,
such a low temperature is almost never reached in the simulated galaxies.

SF and feedback (subgrid physics) are implemented in the code 
as discussed in detail in \citet{Colin+2010} and \citet{Avila-Reese+2011b}. 
Here, we use the same subgrid parameters as in \citet{Avila-Reese+2011b} but 
with better resolution; the size of the cell at the maximum level of
refinement (nominal resolution) is $\approx 60$ pc at $z=0$ and up to $\approx 25$ pc
at the highest redshifts. The actual resolution scale is probably closer to two to four 
times the size of the cell at the maximum level of refinement. Next, we
summarize the subgrid schemes discussed in the above papers.  

SF takes place in all those cells for which 
$T < \Tsf$ and $\rho_g > \rhosf$, where $T$ and $\rho_g$ are the 
temperature and density of the gas, respectively, and $\Tsf$ and $\rhosf$
are the temperature and density threshold, respectively. A stellar
particle of mass $m_* = \epsilon_{\rm SF} m_g$ is placed in a grid cell 
{\it every time} these conditions
are simultaneously satisfied, where $m_g$ is the gas mass in the cell
and $\epsilon_{SF}$ is a parameter that measures the {\it local} efficiency by 
which gas is converted into stars.
No other criteria are imposed. We set $\Tsf = 9000\ \Ke$,  $\nsf = 6\ \cm3$, and
$\epsilon_{\rm SF} = 0.5$ in all the hydrodynamics simulations analyzed in this paper, where
$\nsf$ is the density threshold in hydrogen atoms per cubic
centimeter. 
Observational studies show that the SF rate across entire (local and high-$z$) galaxies 
as well as individual molecular clouds, depends on the mass of very dense gas ($n>10^4$ cm$^{-3}$) 
within molecular cloud complexes, confined to sub-parsec narrow filamentary 
structures and compact cores \citep{Lada+2012,Lada+2010}.
As these authors remark, the key question for understanding what ultimately controls the SF 
is the one related to the local processes that produce the dense and cold gas 
component of the interstellar medium (ISM).
Unfortunately, a density threshold of $n>10^4$ cm$^{-3}$ or more cannot be used in current
cosmological simulations because structures with this density are not resolved. The value used here, 
in combination with the values of other subgrid parameters, is suitable
for our resolution and result in reasonable ISM properties at scales larger than $\sim 100$ pc, 
as well as realistic structural and dynamical properties 
of the whole galaxy \citep[see below; see also][]{Colin+2010,Avila-Reese+2011b}. 
On the other hand, for typical observed column
densities averaged across whole giant molecular clouds ($\langle N\rangle>10^{21}$ cm$^{-2}$), number 
densities corresponding to our resolution at $z=0$ are $n>5$  cm$^{-3}$.

Notice also that, because of our {\it deterministic} SF scheme,
the gas density does not reach values much higher than the SF density threshold. 
Then, it turns out that an $\epsilon_{\rm SF}$
value of about 0.5 gives rise to a conversion of gas-into-stars 
efficiency\footnote{This efficiency can be estimated as the ratio between  the gas infall rate 
and the SFR in the SF cells.} of about a few percent per free-fall time 
(for the chosen density threshold, it is about 20 Myr), 
{\it when} a strong and efficient thermal stellar feedback is present. 
These values are close to the value estimated for 
the giant molecular clouds in the Milky Way by \citet{Krumholz+2007}.

As in \citet{Colin+2010} and \citet{Avila-Reese+2011b}, stellar particles in the simulations here
also inject the energy from supernovae (SNe) and stellar winds in the form of heat into the gas
cells in which they are born. Each star more massive than 8 \msun\ is assumed
to dump into the ISM, instantaneously, $2 \times 10^{51}$ erg in the form
of thermal energy;  $10^{51}$ erg comes from the stellar wind and 
the other $10^{51}$ erg from the SN explosion. Moreover, the star is assumed 
to eject $1.3 \msun$ of metals.

If the resolution is not high enough and/or
$\nsf$ is too high, the cooling time, $t_c$, is comparable or less than the
crossing time, $t_s$, \citep{Stinson+2006,DallaVecchia+2012}, and most of the dumped 
energy is radiated away. It is then a common practice to avoid overcooling 
by delaying the cooling in the star-forming regions
\citep[e.g.,][]{Stinson+2006,Colin+2010, Agertz+2011, Hummels+2012}. We turn off 
the cooling for 40 Myr, after a stellar particle is born, {\it only} in 
the cell where the particle is located. Tests show that the structure of
simulated galaxies is not sensitive to a factor of two variation on the
value of this parameter \citep{Colin+2010}. 
Yet, according to the formulae of the crossing and cooling times in
\cite{DallaVecchia+2012} $t_s \ll t_c$ in the star-forming
cells. These formulae can be applied here 
because the temperature reached by these cells is high, about 
$3 \times 10^7 \Ke$ for the chosen parameters.\footnote{This temperature depends on the total
number of SNe per solar mass, which in turn depend on $\epsilon_{\rm SF}$,
the initial mass function, and the energy injected per SN.} Thus, keeping 
the cooling on or turning it off temporarily is expected to produce similar results. 
In runs not shown here, we see that this is indeed the case.

It should be  said that our density-based deterministic SF and thermal-driven feedback 
implementations, discussed in detail in \citet{Colin+2010} and \citet{Avila-Reese+2011b}, 
have their particularities but, in general agree, with the common schemes applied to this kind
of simulation in order to obtain present-day ``realistic'' galaxies 
\citep[see for a recent review and comparison among several schemes and codes][]{Scannapieco+2012}.

\subsection{The Numerical Simulations}
\label{runs}

We have simulated seven central galaxies with hydrodynamics. In six of the runs, the  \lcdm\ 
cosmological parameters are $\ome = 0.3$, $\omel = 0.7$, $\omeb = 0.045$, and 
$h=0.7$. The fit of the \lcdm\ power spectrum is taken from \citet{Klypin+1997} and is normalized 
to $\sigma_8 = 0.8$, where $\sigma_8$ is the rms amplitude of mass fluctuations 
in 8 \mpch\ spheres. In one more simulation, the cosmological parameters
are $\ome = 0.27$, $\omel = 0.73$, and $\omeb = 0.047$. The power spectrum for 
this latter simulation is the one used to run the ``Bolshoi simulation'' 
\citep{Klypin+2011} with $\sigma_8 = 0.82 $. 

The low-resolution $N$-body simulations from which halos were selected\footnote{
The halos are located using a modified version of the bound density maxima (BDM) halo finder
described in \citet{Klypin+1999}, run only on the dark matter particles.}
are the same used in \citet{Avila-Reese+2011b}. These simulations were conducted
with $128^3$ DM particles in a periodic
box of $10 \mpch$ on a side. As previously mentioned, because our goal is to study
the impact of the halo MAH on the evolutionary properties of
low-mass galaxies, all of our selected halos have about the same mass, around
$2.5\times 10^{10} \msun$ (see Table 1), but different MAHs. 
The halos were chosen to be isolated in the sense that no halo with a comparable or higher mass 
is within a sphere of radius of 1 Mpc. The simulations that include baryons were run 
with high resolution using the ``zoom in'' technique \citep{Klypin+2001}. They
end up with about half a million DM particles inside the high-resolution zone, and
the size of the finest grid cell is 60--30 pc proper. The number of resolution
elements inside the virial radius, \rh, of the halos of the simulated galaxies,
on the other hand, is around 1.5 million. In order to compare the MAHs of
halos with and without baryons, we also ran high-resolution $N$-body-only simulations for 
three of the simulated galaxies. These simulations were run
using the $N$-body version of the ART code \citep{Kravtsov+1997} with comparable resolution
to the hydrodynamic runs.

In ART, the grid is refined recursively as the matter distribution evolves. The runs use a DM 
or gas density criteria to refine. To make sure the dynamics is correctly followed,
we consider a rather aggressive refinement. A cell is thus refined when its mass in DM
exceeds $1.3\ m_p$ or the mass in gas is higher than 1.4$\FbU m_p$, where $m_p$ is 
the mass of the DM particle in the highest resolution region and $\FbU = \omeb/\ome$
is the universal baryon fraction.
In the hydrodynamic simulations presented in this paper, the root grid 
of $128^3$ cubic cells is immediately refined unconditionally to the 
third level, corresponding to an effective 
grid size of $1024^3$.  

Table \ref{table} summarizes the main present-day properties of the seven simulated galaxies/halos.
The halo mass, \mh, is the mass within the virial radius, \rh, defined as the radius that encloses a mean 
density  equal to $\Delta_{\rm vir}$ times the mean density of the universe, where $\Delta_{\rm vir}$ is 
obtained from the spherical top-hat collapse model.
The galaxy properties (\ms, stellar galaxy half-mass radius \re, SFR, etc.) are computed within a sphere of 
$0.1$\rh\ radius. This radius contains most of the stars and cold gas of the simulated central galaxy. 
The contamination of satellites or other substructures at this radius is negligible, and the central galaxies 
hardly extend beyond $0.1$ \rh. 
On the other hand, because the stellar mass density profiles decrease exponentially for most of the runs, 
the galaxy stellar mass does not differ significantly if we measure it at ``aperture" radii smaller than 
0.1\rh\ by factors of 1.5--2.  For example, for the runs Dw1--Dw5, 90\% of the stellar mass at 0.1\rh\ is 
attained at $\sim 0.05$ \rh; for runs Dw6 and Dw7, 90\% of this mass is attained at $\sim 0.085$ \rh\ 
(these late-assembling systems have the most extended stellar mass surface density profiles). 
If we measure \ms\ at half of our ``aperture" radius (0.05 \rh\ instead of 0.1\rh), then the masses would 
be $\approx$ 10\% smaller for runs Dw1--Dw5, and 35\%--40\% smaller for runs Dw6 and Dw7. Because 
observers measure the total luminosity (mass) with different surface brightness (SB) limits, these estimates 
give us an idea of how much could be different in our \ms\ with respect to observational inferences.

Our runs are sorted  and numbered according to their \zh\ value, the redshift at which the given halo
reached one-third of its present-day mass; for similar \zh\ values, 
an earlier mass assembly is determined from the visual inspection of the overall halo MAH. 
In this sense, the halo of run Dw1 forms earlier than the halo of run Dw2 
and so on. {\it This should give us a preliminary idea of how the galaxy properties depend on 
the halo MAH}. We have chosen one-third instead of the standard one-half,
because the MAHs of small halos are such that the fast growth phase, where more stochastic variations
are expected, happens at earlier epochs or at smaller fractions of the present-day mass.

\section{Results} 
\label{results}

\subsection{Properties at $z=0$}
\label{properties}

\begin{figure}
\vspace{8.9cm}
\includegraphics{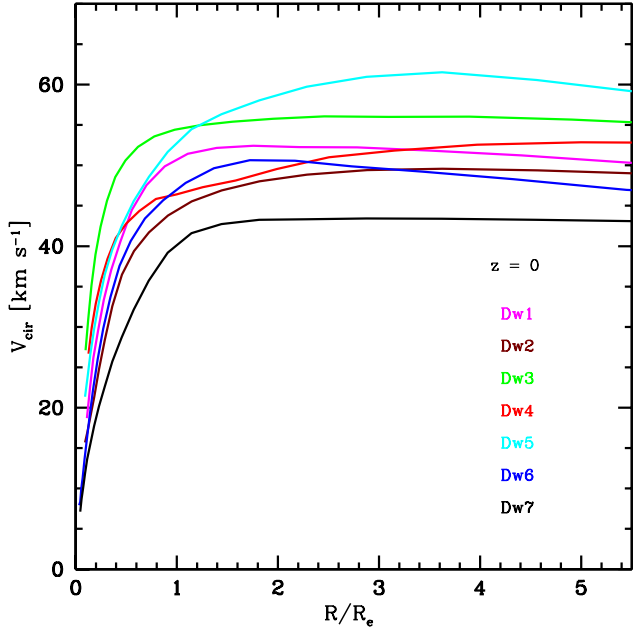}
\caption{
Circular velocity profile, \vcir, at $z=0$ for the different runs with the radius scaled in terms of
the corresponding \re.  
}
\label{vcir}
\end{figure}

\begin{figure}
\vspace{12.1cm}
\includegraphics{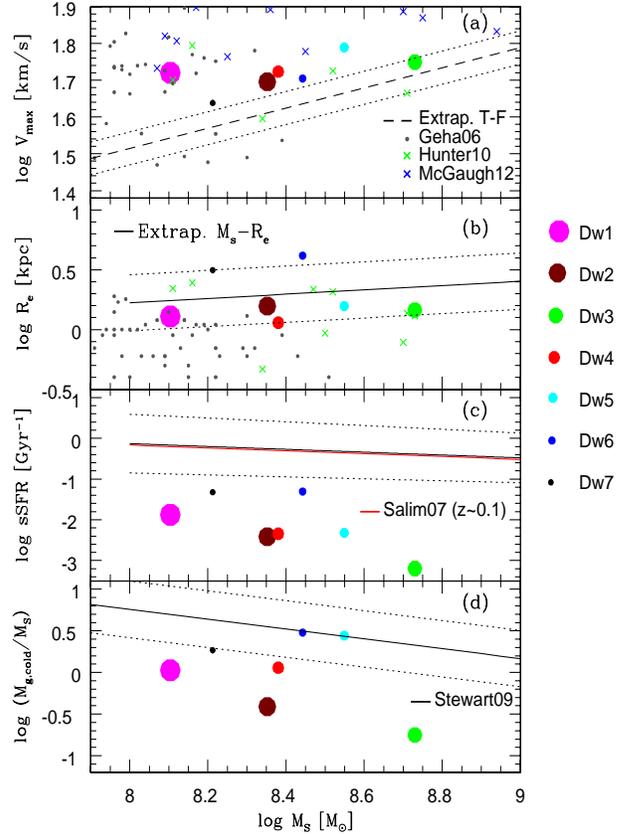}
\caption{
General properties at $z=0$ for the different runs plotted as open circles, where sizes increase as a function of \zh. (a) Maximum circular velocity vs \ms; the dashed line represents the extrapolation of the stellar Tully--Fisher relation as reported in \citet{Avila-Reese+2008}, the small gray points correspond to observations of dwarf galaxies \citep{Geha+2006}, green crosses are $GALEX$ observations of dwarfs reported by \citet{Hunter+2010}, and blue crosses are compiled data by \citet{McGaugh+2012}. (b) Stellar half-mass radius \re\ vs. \ms; the solid line is the extrapolation to low masses of the fit to larger galaxies given by \citep{Dutton+2011}, the dotted lines show the 84th and 16th percentiles of the distribution. (c) sSFR vs. \ms; the solid and dotted lines are the fit, including the scatter to a large star-forming galaxy sample from SDSS, as reported by \citet{Salim+2007}. (d) Gas fraction vs. \ms; the solid line and dotted lines are the analytical fit and its scatter to observations given by Stewart et al. (2009).  
}
\label{correlations}
\end{figure}

The seven present-day isolated halos, where the simulated galaxies form, have 
similar masses (see Table \ref{table}). The resulting galaxies at $z=0$ display a range in stellar
and gas masses that varies by factors between $\sim 4$ and 10, respectively.
The main structural and dynamical properties of the simulated dwarfs are roughly
consistent with those determined for dwarf galaxies \citep[see for a discussion][]{Avila-Reese+2011b}. 
Remarkably, at $z=0$, all have nearly flat circular velocity profiles for radii larger than
$\sim$2--3\re\ (Figure \ref{vcir}). 
In Figure \ref{correlations}, we plot the maximum circular velocity, \vmax, the stellar effective 
radius \re, the \ssfr, and the galaxy cold gas-to-stellar mass ratio, $M_{\rm g,cold}/\ms$,
as a function of  \ms\ for the seven simulated dwarfs (solid colored circles), 
and compare them with some available observational 
information. The sizes of the circles increase with \zh.

According to panel (a), our dwarfs have lower stellar masses
for their  \vmax\ as compared to extrapolations of the (inverse) stellar Tully--Fisher relation
of massive galaxies (dashed line), but are consistent with the observations
of dwarf galaxies (dots). The bending of the Tully--Fisher relation for galaxies
below $\vmax\sim 100$ km $\rm s^{-1}$ has also been found in $N$-body/hydrodynamics 
simulations of a cosmological box by \citet{deRossi+2010}; it is explained by the strong 
effects of the SN-driven outflows in low-mass halos. There is no dependence
of \ms\ or \vmax\ on the halo formation epoch, \zh. Regarding \re, in panel (b), we plot the 
extrapolation to low masses of the \re--\ms\ relation of bigger disk-dominated galaxies 
from Sloan Digital Sky Survey (SDSS) reported in \citet{Dutton+2011}, as well as the measured effective radius
of some late-type dwarf galaxies. The \re\ values of the simulated galaxies are within 
those estimated for the dwarf galaxies, though the scatter
for the latter is large. In several cases, the observed dwarfs are probably satellites. After a galaxy
becomes a satellite, processes such as starvation and ram pressure are expected to quench the 
SF and stop the inside-out growth of the galaxy. Note that the two halos with the latest 
formation epochs, Dw6 and Dw7, have larger \re, by at least a factor of two, than the rest of the 
simulated dwarfs.  As these halos assemble their mass relatively late, they have had more time 
to acquire more angular momentum due to tidal torques during the linear regime. 

According to panels (c) and (d), the simulated dwarfs have present-day \ssfr's and cold gas-to-stellar mass
ratios, $M_{\rm g,cold}/\ms$, smaller than the average ratios estimated for observed galaxies 
of similar stellar masses \citep[see also][]{Avila-Reese+2011b, deRossi+2013}. Note that the
simulated galaxies with lower values of \zh\ (i.e., galaxies that have experienced a relatively late 
assembly) tend to have, on average, higher values of \ssfr\ and $M_{\rm g,cold}/\ms$ ratios, closer 
to observations. Besides, the stellar masses of the simulated dwarfs seem to 
be larger by $\sim 0.5$ dex than what semi-empirical inferences show for halos 
of similar masses (see Section \ref{discussion} for a discussion).

\begin{figure*}
\vspace{14.0cm}
\includegraphics{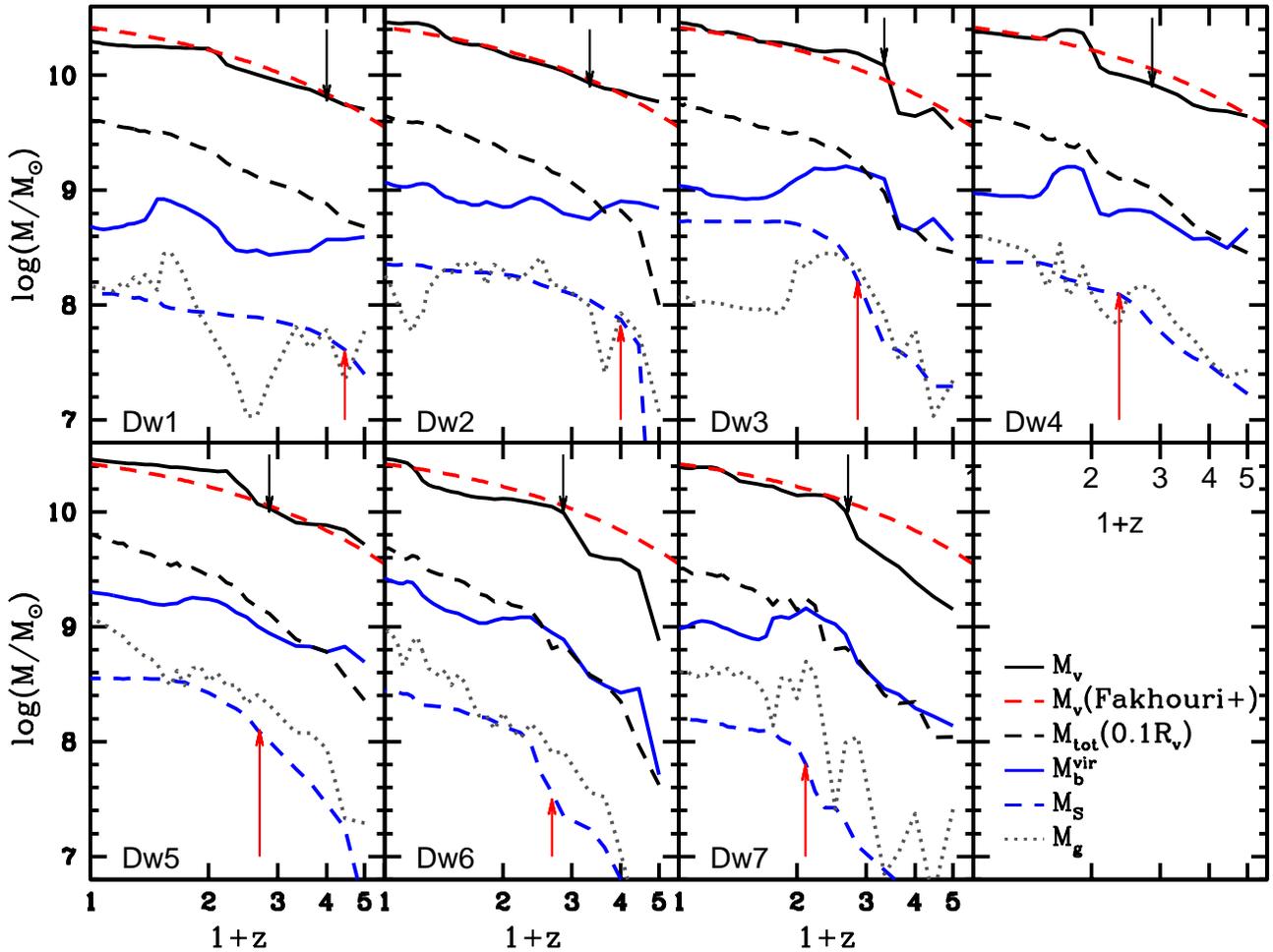}
\caption[\fs\ versus \mh\ ]
{Different mass aggregation histories for the seven simulated galaxies. The solid black (blue) line represents the total (only baryonic) mass inside \rh. The black (blue) dashed line represents the total (stellar) mass inside 0.1\rh, while the gas mass inside 0.1\rh is represented by dotted gray lines. For comparison, in each panel, we plot the average MAH of a halo of log$(\mh/\msun)=10.4$ at $z=0$ as given in \citet{Fakhouri+2010} from an analysis of the Millennium Simulations(red dashed line). The black (red) arrow in each panel indicates the redshift at which 1/3 of the present day \mh\ (\ms) is attained.}
\label{mahs}
\end{figure*}

Most of the gas in the simulated galaxies is cold ($T\leq 10^{4}$ K). In Table \ref{table}, the
$z$=0 mass ratios of cold to total gas in the galaxy, $M_{\rm g,cold}/M_{\rm g}$, are 
reported. For Dw2, this fraction is only about 50\%, while for the rest, it is above $\sim 70\%$. 
On the contrary, most of the gas in the halo, between 0.1\rh and \rh, is hot; in all cases, 
the ratio of hot to total gas is $\sim 99\%$, except for Dw6 which has a huge amount of 
cold gas in the halo. However, in this particular case, we have identified a couple of 
satellites in the halo that contain most of this cold gas.

The stellar structure of the simulated dwarfs varies from simulation to simulation but, in general,
it is composed of a rotating disk and an extended low-angular momentum spheroid;
the total mass surface density profiles roughly follow an exponential law or two exponential
laws, with the outer one being shallower than the inner one. 
The mass contained in the high-angular momentum stellar disk is a fraction of the total stellar mass, 
which ranges from $D/T\sim$0.7--0.6 to $\sim$0.2--0.01 for runs with the latest and earliest halo 
MAHs, respectively. The values of $D/T$ for the seven simulated galaxies are reported in 
Table \ref{table}. It is worth mentioning that the stellar structure of some observed dwarf galaxies 
seems to be dominated in mass by a kind of extended stellar halo with a surface density profile
much flatter than the one of the inner disk
\citep[see, e.g.,][]{deBlok+2006, Barker+2012, Bernard+2012}, likely a product of an early fast 
growth phase of galaxy formation \citep{Stinson+2009}.

\subsection{Mass Assembly Histories}
\label{MAHs}

The main results of the present work are reported in Figure \ref{mahs}, where the 
MAHs of different galaxy/halo components are shown for the seven simulated isolated
dwarfs. The total MAHs (dark + baryonic mass contained within the virial radius \rh; 
solid black lines) are compared with the mean MAH of pure DM halos from the Millenium-2 
simulation which, at $z=0$, end up with the same mass as our runs \citep[$\mh\approx 2.5 \times 10^{10}$ \msun, 
dashed red lines;][]{Fakhouri+2010}. The black arrows indicate the redshift at which one-third of the present-day \mh\ was attained, \zh. For runs Dw1--Dw4, \mh\ attained a third of its 
present value earlier ($\zh\lesssim 2$) with a slower late mass growth
than for runs Dw5--Dw7. The dashed black lines show the total MAHs (dark + baryonic)
but for masses contained within 0.1\rh. The inner mass assembly roughly follows the assembly
of the whole halo, though at earlier epochs ($z>1$), the former is a bit delayed with respect to the latter in 
most runs.

The solid blue lines in Figure \ref{mahs} show the MAHs of the baryon mass (gas + stars) within \rh, \mbv. 
The shape of the virial baryon MAHs partially follows the ones of the total mass (solid black lines), 
with the former being more irregular and with periods of mass decrease, due to gas loss as a result 
of SN-driven outflows.  These periods occur mostly after a major merger has happened (see, e.g., 
the mergers at $z\sim 1.1, 2.2,$ and 2.0 in the runs Dw1, Dw3, and Dw7, respectively, and the 
corresponding later drop of the baryon mass). 
We do not formally construct merger trees, where all of the progenitor subhalos of a descendant halo
at a given time are identified. By major merger, we refer here to an increase in the halo MAH by 50\% (0.176 dex) 
or more between two consecutive snapshots ($\Delta t\sim$300--400 Myr). This increase may come in 
one or several progenitors, although the latter is less probable. In the literature, a major merger is defined 
usually as a merger between two progenitors with mass ratio $M_2/M_1>0.1$ (e.g., \citealp{Fakhouri+2010}).  
Note that at earlier epochs, some of our MAHs 
increase by more than a factor of two, which means that a multiple halo merger event and fast smooth accretion
has occurred.  However, since $z\sim 1$ all of the halos hosting our dwarfs do not increase 
more than a factor of two between two consecutive snapshots.

\begin{figure}
\vspace{7.9cm}
\includegraphics{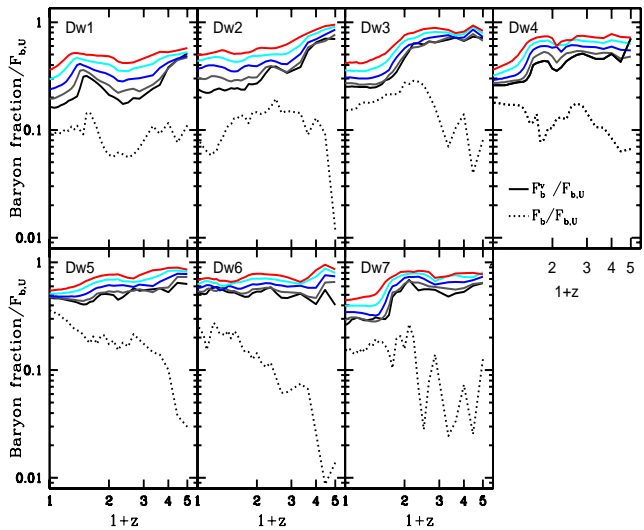}
\caption{
Virial mass baryon fraction, $\Fbv\equiv \mbv/\mh$, as a function of $z$ for all of the runs (solid lines).
From bottom to top, the color lines correspond to baryon-to-total mass ratios inside 1.5, 2, 2.5, and 3 \rh. The
dotted lines are for the galaxy mass baryon fractions, $\Fb\equiv \mb/\mh$. All of the baryon fractions are 
normalized to the universal one, \FbU, defined from the cosmological model used in each run.
}
\label{fbar}
\end{figure}

We define the virial mass baryon fraction as $\Fbv\equiv \mbv/\mh$ and plot this 
fraction as a function of $z$ for all of the runs in Figure \ref{fbar} (solid black lines);
the color lines correspond to baryon-to-total mass ratios inside 1.5, 2, 2.5, and 3 \rh; the dotted line is for the 
commonly discussed \textit{galaxy} mass baryon fraction, $\Fb\equiv \mb/\mh$, where \mb\ is the galaxy 
baryonic mass.  All baryon fractions are normalized to the universal, $\FbU\equiv \omeb/\omem$, defined 
from the cosmological model used in each run.  As can be seen from the plots, \Fbv\ decreases on average 
with time in all runs, except in run Dw6, showing that halos lose more and more baryons (gas) with time. 
Interestingly enough, this behavior extends to regions around the galaxy as far as 3\rh. For the early-assembled 
halos (Dw1--Dw3), the gas loss happens intensively early in the history of their evolution, while for halos assembled later (Dw5--Dw7), the gas loss is less intense overall, ending with higher \Fbv\ values than those 
corresponding to the early-assembled halos.  
In all cases, the virial mass baryon fractions are smaller than the universal one, 
by factors of 1.5--2 at high redshifts, increasing to 2--6 at $z=0$. These factors are much
smaller, especially at high $z$, than those obtained for the galaxy baryon fractions, which shows
that large amounts of gas are actually not in the galaxy but in the halo.\footnote{Some of the baryons in the halo 
are stars but, as it is shown below, in most of the runs the amount of mass in stars that is outside the central 
galaxy is small, especially at low redshifts.} This gas is mostly hot (see Table \ref{table}).   

The dashed blue lines in Figure \ref{mahs} show the galaxy stellar MAHs. 
It is quite remarkable that \textit{the stellar mass assembly of our simulated dwarfs closely follows the halo mass assembly}, at least since $z\sim 1$ for all runs.  
The red arrows indicate the redshift when one-third of the present-day \ms\ is reached, \zg. In general, this 
redshift is close to the corresponding one for total virial mass, \zh\ (black arrows). However, there is a slight 
trend for the early-assembled (late-assembled) halos to assemble their stellar mass earlier (later), 
$\zh \simless \zg$ ($\zh \gtrsim \zg$);
that is, if the halo delays its mass assembly, the corresponding galaxy delays its assembly even more. 
Therefore, {\it an extended halo MAH helps to obtain a galaxy with late stellar mass assembly and higher
SFRs at late epochs}. However, this dependence is actually weak for the galaxies analyzed here.

In Figure \ref{msmh}, we plot the galaxy stellar mass fraction, \Fs, as a function of $z$ for the seven runs 
(solid lines), as well as the halo stellar mass fraction defined as $\Fsv\equiv \msv/\mh$ 
(dashed lines).  For all runs, \Fs\ {\it is almost constant with a value of around $0.01$ since $z=1$}. 
At $z>1$, those halos that have a late assembly history decrease their \Fs\ values with $z$. 
The virial stellar mass fraction, \Fsv, is actually dominated by the central galaxy mass
value; it is only at high redshifts where \Fsv\ is slightly larger than \Fs, thus evidencing 
the presence of some satellites that are probably then accreted by the central galaxy. 
The most remarkable difference is for run Dw6 whose halo MAH grows by jumps (major mergers). It 
is expected, in this case, that the halo and posterior galaxy mergers shall produce those ups and downs in \Fs.  

The dotted lines in Figure \ref{mahs} show the galaxy gas MAHs. Unlike the stellar MAHs, which
always grow, the gas MAHs are irregular, with periods of increase and decrease.  
For runs with late-assembled halos (Dw4--Dw7), the galaxy gas mass, \mg, is mostly
larger than the mass in stars, \ms, while for runs with early halo assembly (Dw1--Dw4), the gas loss
events are stronger, so strong that there are intervals of time for which $\mg < \ms$; the most dramatic
cases occur for Dw3 from $z\sim 1.5$ to $z=0$ and for Dw1 at $1\lesssim z\lesssim 2$. 
The difference between these two cases, however, is that in the latter one, 
the ejected gas is re-accreted later (as evidenced by the much more rapid gas growth rate, as
compared with that of its halo), again increasing \mg, while in the former one,
the gas is lost from the halo. This can be better appreciated in Figure \ref{fgas} below.

\subsection{Gas mass fractions and star formation rate histories}
\label{evolve}

In Figure \ref{fgas}, the ratio between \mg\ and \mb = \mg\ + \ms\ (hereafter \fg, solid lines)
and the halo gas fraction (the amount of gas that it is within 0.1 and 
1 \rh\ relative to the total amount of gas within \rh; dotted lines) are 
plotted as a function of $z$ for the seven runs. It is important to remark that most of 
the gas in the galaxies is cold, while the gas outside the galaxies is mostly hot (see Table \ref{table}).
One can clearly see that in the periods when \fg\ decreases (increases), the gas fraction
outside the galaxy typically increases (decreases), which is clear evidence that \textit{the strong SN-driven outflows 
play a major role in regulating the gas content of the simulated dwarfs}. However, there are some cases
when the outflows are so strong, that the gas is completely lost and the regulation
interrupted. This is the case of Dw3; at $z\sim 2$ (after a major merger), the gas 
fraction in the galaxy, \fg, starts to strongly decrease and it never substantially increases
again because the gas fraction outside the 
galaxy (the reservoir) also decreases, due to the total gas ejection from the halo. 

\begin{figure}
\vspace{7.9cm}
\includegraphics{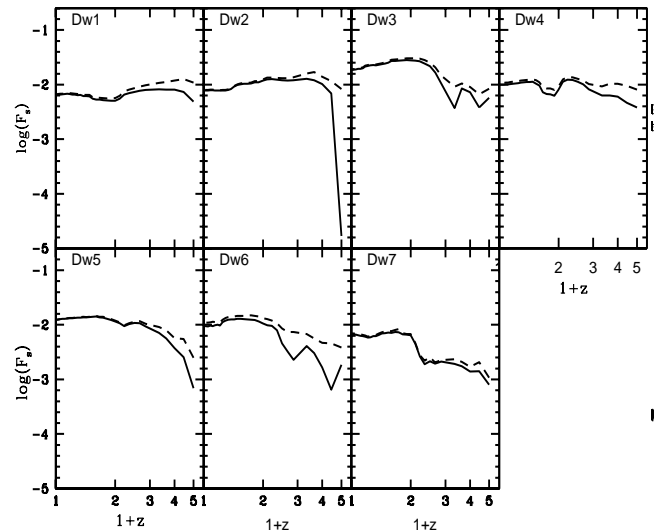}
\caption[\fs\ versus $M_{*}$]
{Evolution of $\Fs=\ms/\mh$. The solid line is for \ms\ inside 0.1\rh, while the dashed line is for \ms\ inside \rh; there is almost no difference because most of the stellar mass is within the central galaxy (0.1\rh). 
}
\label{msmh}
\end{figure}

From Figure \ref{fgas}, one sees that the dwarfs formed in late-assembled halos tend to have higher gas fractions 
(see also Figure \ref{correlations} and Table \ref{table}) and are less episodic than the dwarfs formed in 
early-assembled halos. These latter halos tend to have stronger changes in their virial
baryon fractions, \Fbv\ (Figure \ref{fbar}), which are mostly due to gas ejection from the halo at early epochs,
when the fast mass growth took place; the efficient consumption of gas into stars at early epochs in these halos (runs)
also works in the direction of decreasing \fg. At $z\sim 0$, we then see that \fg\ in the galaxies formed inside early-assembled
halos (Dw1--Dw3, $\fg<0.6$) is lower than in galaxies formed inside late-assembled halos (Dw4--Dw7, $\fg>0.6$). 

In summary, the simulations show some systematic dependences of the gas mass and gas fraction on the halo MAH:
{\it the earlier the halo is assembled, the earlier and more gas is converted into stars and ejected by the SN-driven outflows 
from the galaxy or even from the halo}. When the mass assembly process of the halo is more gradual, as opposed to 
an early fast assembly, the SN-driven outflows seem to be able to regulate the
gas ejection and gas re-accretion processes. These processes, in turn, keep the galaxy with relatively 
high gas fractions, available for SF, so that
a gradual increase in the galaxy stellar mass and fraction (\ms\ and \Fs) with time is observed (see
Figures \ref{mahs} and \ref{msmh}).

\begin{figure}
\vspace{7.9cm}
\includegraphics{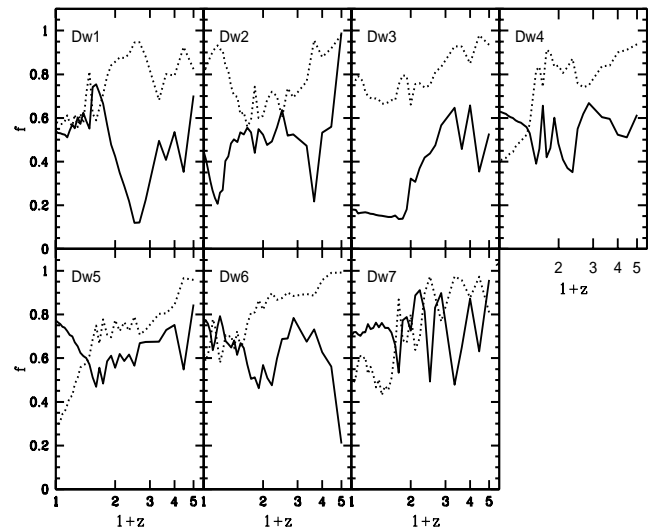}
\caption{
Evolution of the galaxy gas mass fraction, \fg, for the different runs (black solid lines). The dotted line refers to the
mass fraction of gas in the halo (between 0.1\rh\ and 1\rh) with respect to the total amount of gas within \rh.
}
\label{fgas}
\end{figure}

\begin{figure}
\vspace{7.9cm}
\includegraphics{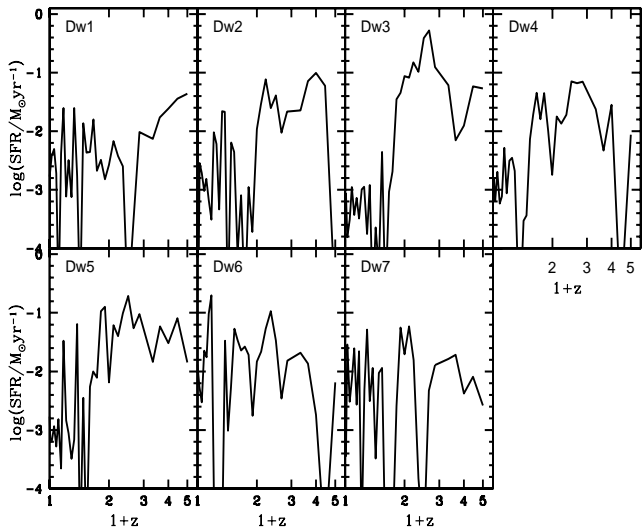}
\caption{
Star formation rate histories for all of the runs. 
}
\label{sfrH}
\end{figure}

The behavior of the galaxy gas mass fraction, \fg, with time is the result of several processes: cosmological gas accretion 
(proportional to the halo MAH) and cooling, transformation of the cold gas into stars, reheating and expansion of 
the gas due to the stellar (mainly SN) feedback, gas ejection, and re-accretion. The interplay of all of these processes 
produces the episodic \mg\ and \fg\ histories seen in Figures \ref{mahs} and \ref{fgas}. Since stars form from the (cold) gas, 
episodic (bursty) SFR histories are expected for our simulated dwarfs. The SFR histories for the seven dwarfs are 
shown in Figure \ref{sfrH}. The SFR is measured as the amount of gas particles promoted to stellar particles inside 
0.1 \rh\ during $\Delta t=100$ Myr at a given $z$. The  episodic character of the SFR histories is clearly reflected in 
the plots. The amplitudes of the burst and quiescent phases in periods of 100 Myr can vary by factors of $\sim$2--10, 
on average, with respect to the current average SFR in periods of 2 Gyr, though in some cases the SFR is 
completely quenched (see Section \ref{inter} for a discussion).
In addition to the strong burstiness, one observes that the average SFH in most of the runs is composed 
of an early ($z\gtrsim 2$) period of high SFRs and then a significant decline at lower redshifts. This 
behavior is not followed by the SFHs of runs Dw6 and Dw7, those with the latest halo assemblies.  
In any case, {\it none of the seven simulated dwarfs can be considered an actively star-forming galaxy at
$z\sim 0$}, contrary to what observations of local isolated dwarfs suggest (see the references in the Introduction).

\section{Effects of baryons on halo mass assembly}
\label{ART}

The effects of the strong gas outflows on the halo MAH of our low-mass halos are not
negligible. There are two main effects that work in the direction of reducing
the total halo (dark + baryonic) mass. The trivial one is the direct loss of baryons from 
the halo, which reduces the total mass. The other effect is dynamical: due
to the mass reduction by baryon losses and a possible expansion effect of the
DM halo produced by the outflows, the gravitational potential of the halo
becomes shallower than in the case with no galactic winds; thus,
the ability of the halo to accrete matter becomes less efficient and the
halo ends up with a lower total mass. 

In order to explore the differences in the virial masses at different epochs, here we compare full 
$N$-body/hydrodynamics (H+ART) simulations with the corresponding $N$-body (ART)-only simulations.  
This exercise is performed for the runs Dw2, Dw5,
and Dw7, for which we have run the corresponding $N$-body-only simulations using the
same initial conditions and comparable (high) resolution. While Dw2 
corresponds to an early-assembled halo with efficient SF and SN-driven outflows
in the remote past, Dw6 and Dw7 correspond to late-assembled halos with respect 
to the average, having a more extended SFH. 

\begin{figure}
\vspace{7.1cm}
\includegraphics{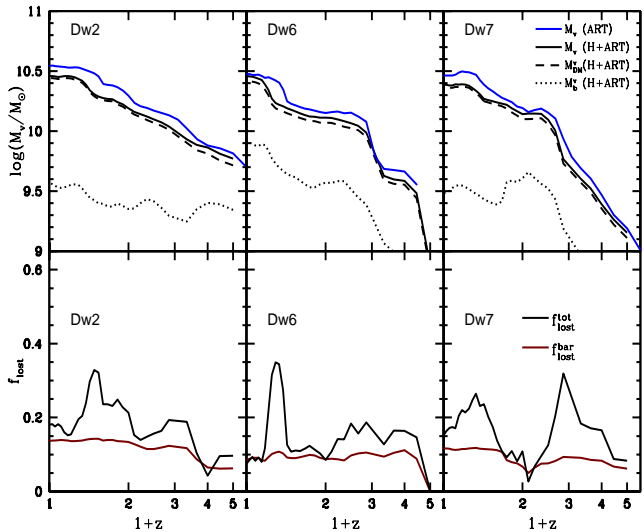}
\caption[sSFR versus $M_{*}$]
{Comparison between the dark matter-only (ART) and the hydrodynamic (H+ART) simulations for the systems
Dw2, Dw5, and Dw7. Upper panels: solid black (blue) line shows the total virial MAH in the H+ART (ART) simulation. 
The black dashed and dotted lines show the dark and baryonic matter components of the H+ART total virial MAHs, respectively;
the latter has been shifted by +0.5 dex. Bottom panels: total and baryonic fractional differences in mass, 
$f_{\rm lost}^{\rm tot}$ and $f_{\rm lost}^{\rm bar}$, as defined in the text.
}
\label{art}
\end{figure}

In the upper panels of Figure \ref{art}, the ART (blue solid line) and the H+ART
(black solid line) total virial MAHs are plotted. As expected, the ART MAHs lie
above the H+ART ones at all times. We also plot the decomposition of the H+ART total virial 
MAHs into DM and baryonic matter, dashed and dotted lines, respectively;
the latter has been shifted by +0.5 dex. 
The fractional difference in mass of the H+ART total virial MAHs with respect to the ART ones, 
$f_{\rm lost}$, is shown in the lower panels with solid lines. This fractional difference
is the result of the above mentioned effects of baryons on the total virial MAHs, and it
is defined as $f_{\rm lost} = [\mh(\rm ART) - \mh(\rm H+ART)]/\mh(\rm ART)$. The strong peaks
seen in the plots should not be interpreted as strong mass differences. In fact, these
peaks appear after mergers. In the H+ART simulations,
mergers occur slightly later than the corresponding mergers in the ART simulations, an 
effect related to the inclusion of the baryon physics 
on the dynamics of the system.
Because of this delay, the fraction $f_{\rm lost}$ increases [$\mh(\rm ART)$ increases
with respect to $\mh(\rm H+ART)$]
until the merger also happens in the corresponding H+ART simulation, and then
returns to, roughly, 
the value it had before the peak.  

Excluding the peaks, we see that the fractional mass difference does not vary significantly 
or systematically with redshift, attaining values between 10\% and 20\% since $z=2$;
that is, {\it the baryonic effect on halos of virial masses $\sim$2--3$\times 10^{10}$ \msun\
decreases their masses (dark + baryons) by factors} 1.1--1.2. Note 
that the larger mass differences are typically reached after the mergers happen.  In order to explore
which is the contribution to the measured mass differences of the simple baryon losses (due
to the SN-driven outflows), in Figure \ref{art}, we plot an estimate of those as:
$f_{\rm lost}^{\rm bar} = [\FbU \mh(\rm ART) - \mbv(\rm H+ART)]/\mh(\rm ART)$ (red line). This fraction
likely overestimates the effect because it is explicitly assumed that halos incorporate
dark and baryonic matter with the same fraction as the universal one.
According to Figure \ref{art}, most of the mass difference is due to the direct gas
losses that result from the SN-driven winds. Note that the maximum value that 
the fractional mass difference can attain due to baryon losses is \FbU, 
in the hypothetical case that all of the baryons are ejected or never reach 
the halo. 

We confirm the results recently reported by \citet{Munshi+2013} and \citet{Sawala+2013},
that at $z=0$, the simulated low-mass galaxies have virial masses smaller 
than the counterpart pure $N$-body simulations. For the halo masses in the range of
our simulations, these authors find differences of 20\%--25\%, which is roughly consistent 
with our results (10\%--18\%). \textit{In addition, we also show that the differences remain 
of the same order since at least $z\sim$2--3}. 
The galaxy outflows are also expected to affect the halo inner mass distribution 
\citep[e.g.,][]{Duffy+2010,Bryan+2013}. Results related to this question will be presented elsewhere.

\subsection{Corrections to the stellar-to-halo mass relation}
\label{ms-mh}
 
The halo/subhalo mass functions obtained in $N$-body cosmological simulations
(pure DM) will change if the effects of baryons are taken into account, especially
at the low-mass side. It is expected that these differences will affect the inferences of 
the \ms--\mh\ relation obtained through statistical approaches as the abundance 
matching technique (AMT) and halo occupation model (HOM) \citep[e.g.,][]{Munshi+2013,Sawala+2013}. 
In the last few years many authors have inferred this relation at $z\sim 0$ and
higher redshifts (see the Introduction for the references), which was used to compare with 
results from simulations and models of galaxy evolution. Of particular interest is
this comparison for low-mass galaxies. In the following, we explore the change in the
\ms--\mh\ relation after taking into account the correction in the halo/subhalo masses as well
as other considerations.

\begin{figure}
\vspace{8.9cm}
\includegraphics{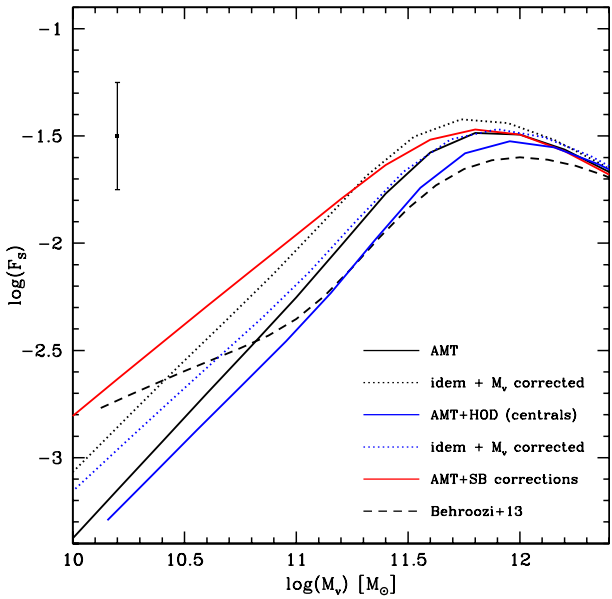}
\caption{
Different cases of semi-empirically inferred \Fs--\ms\ relations: by means of the
AMT (black line) and when the halo/subhalo mass function is corrected for 
the effects of baryons (dotted line);  by means of the AMT+HOD, only for
central galaxies/distinct halos (blue line) and when the correction for the effects
of baryons is introduced (blue dotted line); again, by means of the AMT but
correcting the GSMF for the (uncertain) SB issues (red solid line); see the
text for details. The dashed line reproduces the AMT inferences by
Behroozi et al. (2013). The error bar indicates the typical systematical
uncertainty (mainly due to the \ms\ determination) of these inferences.
}
\label{AMT}
\end{figure}

In order to statistically infer the \ms--\mh\ relation down to low masses, we construct a galaxy stellar 
mass function (GSMF) similar to the one reported in \citet{Baldry+2008}.  For this, we use the catalog
used by these authors, namely, the SDSS DR4 version of the New York University Value-Added Galaxy
Catalog \citep{Blanton+2005a,Blanton+2005b}, with their $V_{\rm max}$ volume correction and the stellar 
mass calculated from the $g$ and $i$ bands according to the \citet{Bell+2003b} mass-to-luminosity ratios. 
The GSMF is close to the one reported by these authors but in the $10^9$ to $3\times 10^{10}$ \msun\ 
interval, our GSMF is slightly higher and less curved than in Baldry et al. (2008).
In Figure \ref{AMT} we plot the \Fs--\mh\ relation obtained for this GSMF by applying
the AMT as in \citet[][black solid line]{Rodriguez-Puebla+2012}. The error bar in the panel 
indicates the typical uncertainty in the \Fs\ determination due to systematical uncertainties,
mainly the one in the stellar mass. 
We further take into account the effect of baryons on the halo mass discussed above by using 
the correction on \mh\ given in \citet{Sawala+2013}.  As shown above, our results 
are consistent with this correction, at least at the masses studied here. After applying this correction to the
halo/subhalo mass function and applying the AMT, we obtain
the \Fs--\mh\ relation plotted in Figure \ref{AMT} with the dotted black line. 

The common AMT inferences do not make a difference between the \Fs--\mh\ relation of
central and satellite galaxies. As shown in Rodr\'iguez-Puebla et al. (2012, 2013), they are actually
different. In order to constrain both relations, these authors used the AMT combined with 
the HOD model; the latter requires information about the observed two-point correlation
function. Following \citet{Rodriguez-Puebla+2013}, we calculate the \Fs--\mh\ relation
separately for centrals/halos and satellites/subhalos for the same total
GSMF discussed above; the result for \textit{centrals/halos}, 
is plotted in Figure \ref{AMT} as the blue solid line, while the dotted blue line is for
the case where \mh\ in the halo mass function is corrected for the effects of baryons. The
latter relation will be used in Figure \ref{mbar-mh} of Section \ref{outflows} for comparison 
with our simulated central dwarfs.

The short-dashed line in Figure \ref{AMT} corresponds to the AMT result by Behroozi et al. (2013),
who used a combined GSMF: from \citet{Moustakas+2013} for large masses and from 
Baldry et al. (2008) for low masses. For the definition of halo mass, instead of the present-day 
halo mass or the mass at the accretion epoch in the case of subhalos, they use the maximum 
mass a halo/subhalo ever had (peak mass). Therefore, their halo mass function is expected 
to be slightly higher than the one used by us. This is partially why their \Fs\ values for 
$\mh\gtrsim 10^{11}$ \msun\ are slightly below our corresponding AMT result (solid black line).  
The other reason is due to the small difference between our and the original Baldry et al. (2008) 
GSMF used by these authors (see above).
The strong flattening of the \Fs--\mh\ relation at masses below $\sim 10^{11}$ \msun\ in 
Behroozi et al. (2013) is probably due to a correction for SB 
incompleteness applied by them but not described in the paper. 
In order to explore this question, we apply a correction to our GSMF for this incompleteness 
and for the SB--magnitude correlation by following the recipes given in \citet{Blanton+2005a}. 
The obtained \Fs--\mh\ relation with the AMT lies above and becomes shallower at the low-mass side 
(red solid line) than the case without this correction (black solid line); 
at $\mh=3\times 10^{10}$ \msun, \Fs\ is already 0.5 dex higher after the (uncertain) SB corrections.  
However, we do not reproduce the strong bending reported in Behroozi et al. (2013).
It seems that their SB corrections are stronger at low masses than those suggested 
by \citet{Blanton+2005a}.

We conclude that (1) correcting the halo/subhalo mass function by the effects of baryons 
does not significantly affect the \Fs--\mh\ relation, at least down to $\mh\sim 10^{10}$ \msun;
(2) for central galaxies, \Fs\ is lower than the average case (conversely, for satellites, it is higher;
see \citet{Rodriguez-Puebla+2012}; (3) if the SB corrections are strong, the \Fs--\mh\ relation could increase significantly
at lower masses in such a way that the low-mass
end of the GSMF would become very steep (see also Rodr\'iguez-Puebla et al. 2012; \citealp{Sawala+2013}).

\section{Discussion}
\label{discussion}

\subsection{SF-driven outflows or delayed SF?}
\label{outflows}

Despite the fact that the seven simulated dwarfs show structural and dynamical
properties roughly consistent with observations, they have too low \ssfr's and
gas fractions as compared to the values estimated for local dwarfs (central or satellites), 
and presumably too high stellar masses as well (see below), a consequence of an 
efficient gas transformation into stars at early epochs. Overall, our analysis has shown 
that {\it the stellar mass growth of the dwarfs closely follows the mass growth of their halos}. 
In this sense, the dwarfs formed in those halos with a late mass assembly tend to 
have a late stellar mass assembly as well, with higher present-day SFR's and gas fractions. 
Thus, dwarfs formed in halos that assemble their masses later than the average have 
evolutionary signatures closer to observations; though, even in these cases, the simulated 
dwarfs are not as {\it active} and gas rich as observations suggest 
\citep[see also][]{Colin+2010,Sawala+2011,Avila-Reese+2011b,deRossi+2013}.

Gas outflows or supergalactic winds driven by stellar feedback (mainly SNe) have been commonly 
invoked as the mechanisms able to lower the stellar-to-halo mass ratio, \Fs, and obtain the flattening of the faint-end of the GSMF. The effects of the 
SF-driven feedback (injected to the ISM as energy and/or momentum) are also expected to influence 
the thermal and hydrodynamic properties of the gas in such a way that the SFR history is affected 
by the feedback. Our hydrodynamical simulations include an efficient prescription for thermal feedback from 
stars and SNe (see Section \ref{the_code}).

According to the results presented in Section \ref{results}, the gas outflows in the simulations are strong, removing 
large fractions of baryons not only from the galaxies, but also from the halo, and producing episodic SFHs. 
However, in spite of it, our results do not agree with observational inferences as mentioned above. 
The question is {\it whether stronger SF-driven outflows (higher mass loading factors) should be allowed in order to 
solve the issues of the simulated galaxies or there is no more room for strong outflows in the simulations}. 

\begin{figure}
\vspace{8.9cm}
\includegraphics{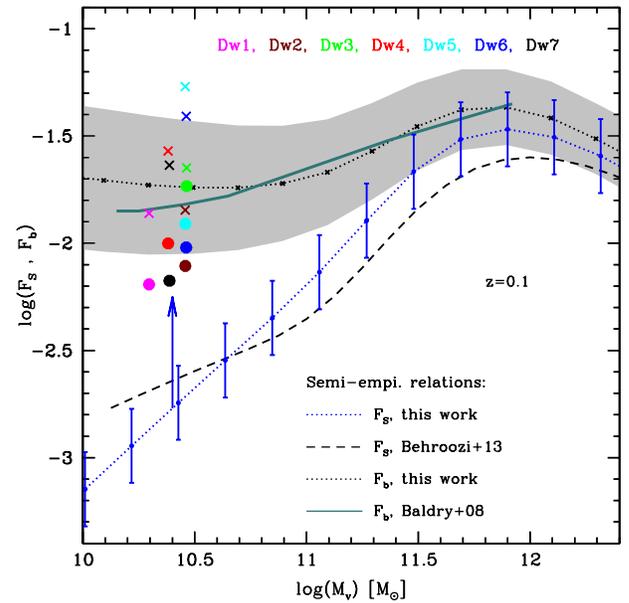}
\caption{
Stellar-to-halo (filled circles) and baryonic-to-halo (crosses) mass ratios
vs. \mh\ for the seven simulated dwarfs. The blue dotted line with error bars corresponds to the \Fs--\mh\ relation for central
galaxies/distinct halos and its intrinsic scatter plotted in Figure \ref{AMT}. The arrow indicates the factor by which \Fs\ increases at 
log(\mh/\msun) = 10.4 if the (uncertain) corrections to the GSMF for SB issues are introduced (see Section \ref{ms-mh}). The dashed line corresponds
to the \Fs--\mh\ relation as inferred semi-empirically by 
Behroozi et al. (2013). Our inferences of the corresponding \Fb--\mh\ relation and its 
$1\sigma$ scatter are represented by the dotted line and the gray shaded area 
(see the text for a description of how this is calculated). We also reproduce the mean 
\Fb--\mh\ relation calculated by Baldry et al. (2008; green thick line).
}
\label{mbar-mh}
\end{figure}

In Figure \ref{mbar-mh}, we plot the $z=0$ stellar-to-halo (filled circles) and baryonic-to-halo 
(crosses) mass ratios versus the corresponding virial masses for the seven simulated central dwarfs. 
The blue dotted line with error bars corresponds to the \Fs--\mh\ relation for central
galaxies/distinct halos and its intrinsic scatter as inferred in Section \ref{ms-mh} (Figure \ref{AMT}) 
by means of the AMT+HOD formalism, \textit{taking into account the correction to the halo masses
due to the effects of baryons}. The arrow indicates the factor by which \Fs\ increases at 
log(\mh/\msun) = 10.4 if the (uncertain) corrections to the GSMF for SB issues are introduced 
(see section \ref{ms-mh}). The dashed line is for the average (central and satellites) relation reported 
in Behroozi et al. (2013). The simulated dwarfs are, on average, $\sim 0.7$ dex above the 
semi-empirical inferences.

What about the baryonic-to-halo mass ratio, \Fb, versus \mh ? In order to calculate this relation,
we add (see Section \ref{ms-mh}) information on the
gas content to the galaxy catalog used here, thus constructing the corresponding galaxy baryonic mass function (GBMF), 
and applying the same AMT+HOD formalism mentioned above. For the gas 
content, we use the empirical \mg--\ms\ relation given in \citet{Stewart+2009}, including its scatter. 
The result obtained for central galaxies/distinct halos is plotted with the dotted line and gray 
area ($1\sigma$ intrinsic scatter) in Figure \ref{mbar-mh}. We also plot the inference by Baldry et
al. (2008; green thick line), who used their GSMF, the empirical \ms--metallicity relation, and
a  model to infer from it the \fg--\ms\ relation, for obtaining the GBMF. By abundance matching 
this function with the CDM halo mass function, they calculated the mean \Fb--\mh\ 
relation that we reproduce in Figure \ref{mbar-mh} (green thick line).  For a given halo mass, 
the baryonic mass ratios, \Fb, of our dwarfs are close to those inferred semi-empirically, unlike 
what happens with the stellar mass ratios, \Fs. This result is 
consistent with the fact that the gas fractions of the simulated dwarfs are lower, on average, than 
what observations show (Figure \ref{correlations}).

The comparison shown in Figure \ref{mbar-mh} suggests that {\it there is not much room for 
more efficient SN-driven outflows (with higher mass loading factors) than those obtained in the
simulations}; otherwise, the baryonic masses of the simulated dwarfs would be too low with 
respect to the semi-empirical inferences.  Besides, more efficient outflows are not expected to 
help in keeping the SF at low redshifts active, as observations suggest for most low-mass galaxies, especially the isolated ones \citep[cf.][]{Salim+2007,Geha+2012,Pacifici+2013}.
Instead, more gas would be ejected from the galaxy and halo, making a
later re-accretion very unlikely (necessary to fuel SF) given the small and almost non-increasing gravitational potential 
of these low-mass halos \citep[see][for a discussion]{Firmani+2010a}.  

Therefore, the avenue for improving the simulations of low-mass galaxies,
rather than increasing the strength of the SF-driven ejective feedback, should take into account processes that lower the SF efficiency at early epochs and delay the 
stellar mass assembly of the dwarfs with respect to their halo. This may be attained, for example, 
by (1) including the momentum transfer to the gas by the radiation 
field from young massive stars (radiation pressure) and the heating by local photoionization in the stellar feedback recipe
\citep[preventive feedback; e.g.,][]{Murray+2005,Brook+2012,Hopkins+2012,Wise+2012,
Agertz+2013,Ceverino+2013}; or (2) by introducing an H$_2$-based
SF scheme instead of a gas density threshold \citep[][]{Krumholz+2011,Kuhlen+2012,Christensen+2012,Munshi+2013,Thompson+2014}. It is known that H$_2$ formation depends on the gas metallicity; 
therefore, the SF should be less efficient in the past for the low--metallicity, low--mass 
galaxies. 

\citet{Christensen+2012} simulated a dwarf galaxy of halo mass similar to ours, where
SF is triggered only in those regions where H$_2$ was formed (a scheme of H$_2$ metallicity--dependent
formation was implemented). They compared their results with a similar simulation but with the
usual SF density threshold recipe with $n_{\rm SF}=100$ $\rm cm^{-3}$. They find that at high $z$, when the 
metallicities are low, both simulations show similar results (see their Figure 8), suggesting that  
a high $n_{\rm SF}$ value could emulate the H$_2$-driven SF implementation.
We experimented with a $n_{\rm SF}$ value of 100 cm$^{-3}$ in our Dw1 and Dw3 galaxies. 
The obtained dwarfs are unrealistic, with strongly peaked circular velocity profiles and stellar and gas
masses at $z=1$, larger than those presented in Section 3 \citep[see also][]{Colin+2010}. 
With about the same energy injection, specially at high redshift, the (strong) SN feedback is now unable to disperse 
and blow the gas out of the high-density lumps (comprised of the cells where SF proceeds 
and those around) and to avoid the SF runaway, as well as the formation of massive clumps 
that migrate to the center, making the galaxies very concentrated.

\subsection{Where Are the baryons?}
\label{baryons}

\begin{figure}
\vspace{8.9cm}
\includegraphics{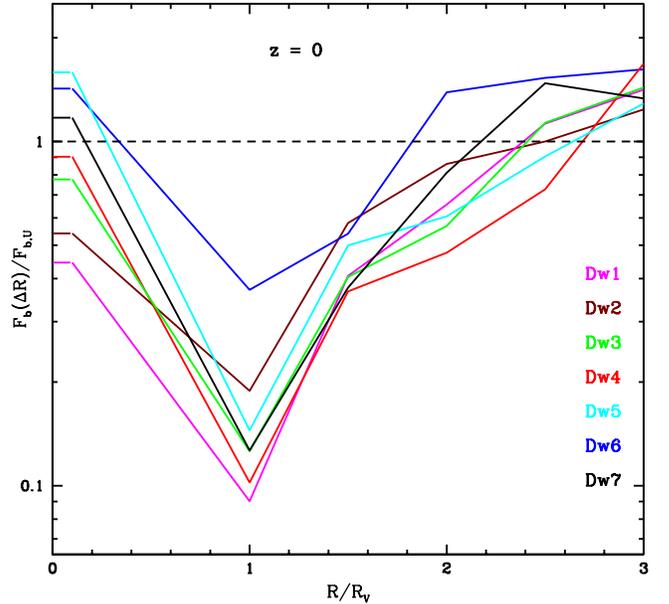}
\caption{
Different spatial baryon mass fractions (at $z=0$), $F_{b}(\Delta R)\equiv \mb(\Delta R)/ \mh(\Delta R)$, 
where $\Delta R$ are radius intervals in unities of \rh\ between 0 and 0.1,
0.1 and 1, 1 and 1.5, 1.5 and 2, 2 and 2.5, and 2.5 and 3.0.
}
\label{fbaryons}
\end{figure}

Figure \ref{fbar} shows that the total baryon mass fraction within the virial radius, \Fbv\ 
(solid black lines), is significantly lower than the universal baryon fraction, \FbU,
specially at lower redshifts. This fraction at radii larger than \rh\ 
is still smaller than \FbU\ (solid color lines, for radii up to 1.5, 2.0, 2.5, and 3.0 \rh).
In Figure \ref{fbaryons}, we show the spatial baryon mass fractions at different radii 
(spherical shells), $F_{b}(\Delta R)\equiv \mb(\Delta R)/ \mh(\Delta R)$,
in units of \rh\ for the seven simulations. These fractions are evaluated at 
$z=0$ (according to Figure \ref{fbar}, the results should be qualitatively similar at higher redshifts). 
The radii, in units of \rh, at which $\Delta R$ is defined are between 0 and 0.1, 
0.1 and 1, 1 and 1.5, 1.5 and 2, 2 and 2.5, and 2.5 and 3.0. 
The baryon fractions are normalized to the universal one.  

From Figure \ref{fbaryons}, we identify four regimes for the baryon mass distribution with respect to the total one: 

1. The baryon-to-total mass ratios inside 0.1 \rh\ tend to reach high values; this is the place where most
baryons settle to form the galaxy. 
However, notice that, even in this region of high concentration, the baryon fraction is lower than the universal one for four 
of the dwarfs; the three for which the local baryon fraction is slightly higher than \FbU, are those that assemble 
their mass later (as already
discussed, these galaxies suffer less early gas loss). Thus, at 0.1 \rh, DM dominates locally 
in all of the simulated dwarfs, something that agrees with dynamical studies within the optical radii of observed dwarfs.  

2. The baryon-to-total mass ratios in the shell with radii from 0.1 \rh\ to 1 \rh\ are between 0.1 and 0.5 \FbU; i.e., these are
baryon-deficient regions because both a fraction of the baryons flowed to the center to form the galaxy
and another larger fraction was blown away due to the galaxy feedback. 
As mentioned in Section \ref{MAHs} (see Figure \ref{msmh}), most baryons in the halo are 
in the form of hot gas. The virial temperature corresponding to halos of circular velocities $ \simless 60$ km $\rm s^{-1}$ is actually very low; 
hence the halo gas in the simulations is hot due to the galaxy SN-driven feedback. Most of this gas is 
actually outflowing material. 

3. The baryon-to-total mass ratios at radii larger than the virial radius are larger than those inside it but still lower than 
\FbU, at least up to 2--3 \rh. This means that the effects of the galaxy feedback extend out to these large 
radii, and gas is still outflowing here. Notice that 2--3 \rh\ correspond to 100--150 kpc $\rm h^{-1}$ in physical scales.

4. Only at radii as large as 2--3 \rh, the baryon-to-total mass ratios become close to the universal baryon fraction,
and at still larger radii, there is even an excess of baryons with respect to the universal average. 
This excess can be explained as the accumulation of baryons swept out by the galaxy outflows.

\subsection{Episodic star formation}
\label{inter}

\begin{figure}
\vspace{8.9cm}
\includegraphics{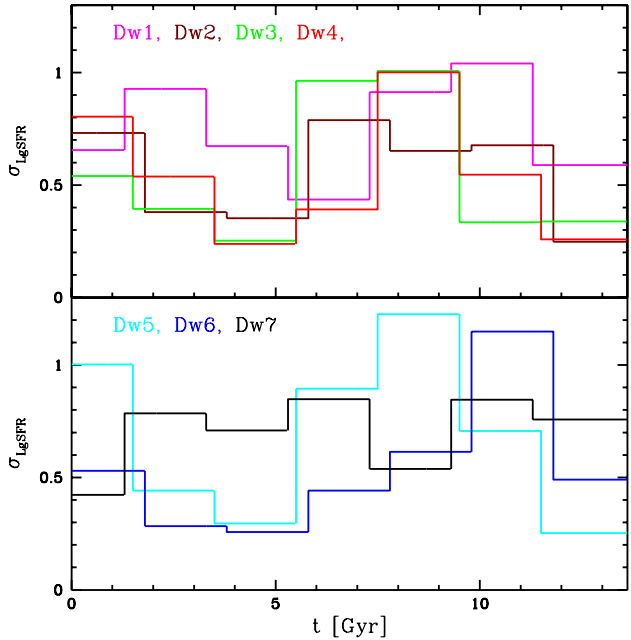}
\caption{
Logarithmic 
standard deviation of the SFR measures in periods of $\Delta t=100$ Myr with respect to the mean SFR in a period
of 2 Gyr, $\sigma_{\rm LgSFR}$, as a function of cosmic time for the seven runs divided in the two panels. 
An ``archeological'' SFR history is used to compute $\sigma_{\rm LgSFR}$.
}
\label{bursty}
\end{figure}

According to Figure \ref{sfrH}, the history of the SFR of the simulated dwarfs is episodic, with strong ups and downs, 
even with periods of time in which it is completely quenched. 
In an attempt to quantify the average amplitude of the bursty
SFR behavior, we compute the logarithmic 
standard deviation of the SFR (measured every $\Delta t=100$ Myr) with respect to its mean in a period
of 2 Gyr, $\sigma_{\rm LgSFR}$.  The results for the seven dwarfs are plotted in Figure \ref{bursty}. For this exercise, 
we use the stellar age histogram as a proxy for the SFH; as we keep record of the birth time of each 
stellar particle, we use this to compute the amount of mass in stars inside the galaxy in time bins 
of 100 Myr width, from the oldest to the youngest stellar particle.\footnote{We have decided to 
use 100 Myr to be consistent with the analysis done in Figure \ref{sfrH}, where 
the ``instantaneous'' SFRs were computed using this period of time. However, we have checked that 
the global results regarding the measured SFRs do not change significantly if $\Delta t$ is assumed to be larger 
(200 Myr) or smaller (50 Myr) than our assumed value.} 

From Figure \ref{bursty}, we observe that the logarithmic amplitude of the 100 Myr episodes of the SFR oscillates
between $\sigma_{\rm LgSFR}=\pm 0.3$ and $\pm 1$ dex for all of the runs and at different cosmic epochs. 
There is not a monotonic trend of $\sigma_{\rm LgSFR}$, calculated in periods of 2 Gyr, with cosmic time.
For most runs, the degree of burstiness attains a maximum at early epochs, then decreases, and
after 5--6 Gyr, it increases again, just to show a fall in the last bin. This behavior is more pronounced for galaxies in early
assembled halos. The fact that the SFR changes in 100 Myr periods by factors of 2--10, on average,
in our simulated galaxies is partially explained by the effects of stellar feedback, which likely acts at the level of the whole galaxy, heating and ejecting the gas from it. As a consequence,
the global SFR is decreased or quenched until gas infalls and cools again. The bursty SFH
can also be due to the small number of SF regions in the dwarfs; since these regions are intrinsically
stochastic, a low number of them also implies a somewhat stochastic global SFR. 
In general, our results are qualitatively similar to the ones found by other authors, 
which also report a bursty 
behavior in the SFR of dwarf simulated galaxies \citep[e.g.,][]{Stinson+2007,Teyssier+2013}. In 
\cite{Teyssier+2013}, the extreme fluctuations in the SFR reach amplitudes of a factor of $\sim 10$. 

It is difficult to evaluate whether the bursty nature of the SFHs of our
simulated dwarfs is consistent with the SFHs of  
observed dwarf galaxies because, from the observational point of view,
the number and amplitude of the SF episodes are not well constrained.  
Only limited and/or indirect inferences have been obtained.
From observations of the Local Group dwarf galaxies, \citet{Mateo+1998} found that the most 
recent SF episodes last from 10 to 500 Myr in both irregular and early-type dwarfs.
On the other hand, based on the 11HUGS sample, and by assuming that bursts occur with equal probability in 
all the sample galaxies, \citet{Lee+2009} found that the SFR in the burst mode is $\sim 4$ 
times greater than in the quiescent mode. They also estimated that the bursts, assumed to last $\sim$ 100 Myr
\citep{Weisz+2008}, occur every 1--2 Gyr, and the fraction of stars formed in the bursts is, on average, $23\%$.
For a sample of eighteen nearby starburst dwarfs, \citet{McQuinn+2010} reported that their recent 
SFRs are more than two times higher than the SFR averaged over  the past 6 Gyr.

In a detailed study of 60 dwarfs from the ACS nearby galaxy survey treasury, \citet{Weisz+2011b} 
showed that SFHs are so complex that they cannot be explained by simple SFH models, like 
single bursts, constant SFRs, or exponentially declining SFRs.
The above mentioned observational estimates confirm that dwarf galaxies have an episodic SFH,
in general \citep[see also][for an extensive discussion]{Bauer+2013}. However, the amplitude and frequency 
of the SFR episodes seem to be lower than those measured in our simulated dwarfs. More 
observational work is necessary in order to quantitatively evaluate how episodic the SFHs of dwarf galaxies are. If the conclusion that the simulated
dwarfs have SFHs similar to or burstier than the ones inferred for observed galaxies holds, {\it this then indicates 
that the intensity of the SN-driven global outflows in the simulations cannot be
stronger}; otherwise, the SFHs would show even more frequent and more intense
galactic SFR episodes.

\subsection{Numerical resolution}
\label{resolution}

In order to determine the effect of resolution in our results, we ran galaxies Dw1 and Dw3 with less resolution: 
the cell size at the maximum level of refinement is twice as large ($\sim 109$ pc), 
and the mass of the DM particle $m_{p}$ is eight times 
higher than in the runs presented above. The two runs with lower resolution, Dw1$_{\rm lres}$ and Dw3$_{\rm lres}$,
end with galaxy properties and evolutionary trends similar to runs Dw1 and Dw3, respectively, showing that 
we have found resolution convergence, at least in regards to the aspects studied here.  
For example, the stellar mass fractions at $z=0$ in Dw1$_{\rm lres}$ and Dw3$_{\rm lres}$ are log(\Fs)=$-2.2$ and --1.7,
respectively, very similar to those obtained in the runs Dw1 and Dw3; the gas fractions also do not vary by
more than 5\% between the simulations with the two resolutions; the circular velocity profiles are again flat at large 
radii, with small changes in its maximum, \vmax(Dw1$_{\rm lres}$, Dw3$_{\rm lres}$) = (55.5, 54.6) km s$^{-1}$ versus 
\vmax(Dw1, Dw3) = (52.0, 56.1) km s$^{-1}$. Moreover, the halo, baryonic, and stellar MAHs look similar.

\section{Conclusions}

High-resolution zoom simulations of seven central dwarf galaxies formed in isolated halos, 
which today attain the same virial (dark + baryonic) mass, log(\mh/\msun)$\approx 10.4$, were 
performed with the $N$-body/hydrodynamics code ART, including standard prescriptions for SF and 
SF-driven thermal feedback. All simulated dwarfs have nearly 
flat rotation curves, with some of them also having a 
significant stellar disk component. They roughly agree with observations in the \vmax--\ms\ and \re--\ms\ 
relations. However, their \ssfr s and gas fractions are lower than the observational determinations, 
and their stellar masses are too high as compared to 
semi-empirical inferences of \Fs=\ms/\mh; on the other hand, 
their baryonic masses (stars + gas) seem to 
be in agreement with these kind of inferences. The seven runs were classified 
according to their halo MAHs, from those that assemble their masses, on average, earlier to those that do it later 
(from Dw1 to Dw7, respectively). We then explored how the properties and mass assembly of the different 
galaxy/halo components of the dwarf systems depend on the halo MAHs. Our main conclusions are as follows.

\begin{itemize}

\item Stellar mass assembly of the central dwarfs closely follows their halo mass assembly; 
the ratio \Fs\ is $\approx 0.01$ and constant since $z \sim 1$ for all of the simulated galaxies, and
at higher redshifts, the dwarfs formed in the late-assembled halos tend to have even smaller \Fs\ values. 
The baryons within the virial radius roughly follow the assembly of the whole halo but there are
periods when baryons are lost due to the SN-driven outflows; as a result, the  
\Fbv=\mbv/\mh\ ratio systematically decreases with time, from values 1.5--2 times smaller 
than the universal baryon fraction at $z\sim 4$ to values 2--6 times smaller at $z=0$. For 
the early-assembled halos, the SN-driven gas loss happens intensively early
in the evolution, while for 
halos that assemble their mass later, the gas loss is less intense overall, ending these systems with higher 
\Fbv\ values than those assembled earlier. The mentioned behaviors of the baryon content 
with respect to total mass extend roughly up to 3\rh\ (100--150 kpc $\rm h^{-1}$ from the galaxy in 
physical scales), showing a high efficiency of the gas outflows in the simulations. 

\item Assembly histories of the gas of the dwarfs are episodic, with periods of increase and decrease. The 
dwarfs formed in late-assembled halos have higher gas fractions, \fg=\mg/\mb, than those formed
in early-assembled halos; the early SF-driven outflows remove more gas from the latter than the former.
When the galaxy \fg\ decreases, one typically sees an increase in the gas fraction in the halo (this gas
is much hotter, while the gas in the galaxy is dominantly cold). This shows that the strong SN-driven 
outflows play a major role in regulating the gas content---and therefore the SFR---of the simulated dwarfs,
at least for those in which the corresponding halo assembling happens more gradually. 
However, there are also periods, mostly in the early-assembled 
halos, when the gas fractions in both the galaxy and halo decrease, since the gas is 
completely expelled from the halo.   

\item SFHs of the simulated dwarfs are quite episodic with average variations of the SFR 
(measured every 100 Myr) of factors 2--10 with respect to the mean, measured in
periods of 2 Gyr. The average 
SFH in most of the runs is composed of an early ($z\gtrsim 2$) period of high SFRs and then a 
significant decline at lower redshifts; the exceptions are the two dwarfs formed in the latest-assembled halos (Dw6 and Dw7). However, even in these cases, the SFHs do not show the late
active SF regime of observed isolated dwarfs.

\item The effect of baryons on the total virial mass of the simulated halos is to reduce
it by 10\%--20\% with respect to that obtained in the $N$-body only simulations; this effect
is seen from at least since $z\sim 2$. 
Most of the difference is caused by the loss of baryons from the halo due to the SN-driven outflows. 
A smaller contribution comes from the gravitational potential being less
deep, due to gas loss, and the halos are less capable of accreting mass. The abundance 
matching and halo occupation model carried out with a corrected halo/subhalo mass function
by the effects of baryons gives a slightly higher \Fs--\ms\ relation at lower masses than in the
case when no correction is applied.

\end{itemize}

In summary, we conclude that the ``cosmological'' halo MAHs have a non-negligible
influence on stellar 
and baryonic mass assembly of the dwarfs that form in their centers. In spite of the strong 
SF-driven feedback effects (mainly the SN-driven outflows), the mass assembly of galaxies roughly
follows the one of their halos. Since, in the hierarchical \lcdm\ scenario, low-mass halos 
are assembled early, then low-mass galaxies are expected to have an early stellar 
mass assembly and small \ssfr s at late epochs. Our results indeed show that the simulated galaxies,
even those that form in late-assembled halos, have 
lower \ssfr s and gas contents, and larger stellar masses for their halo masses than 
observations and semi-empirical inferences show. Yet, the baryonic masses of the simulated 
dwarfs seem to agree with the semi-empirical inferences. 
Thus, rather than further increasing the strength of the ejective SN-driven feedback, possible avenues to solve the issues of
simulated low-mass galaxies are to introduce subgrid processes that delay the atomic
gas transformation into molecular gas, and/or to take into account  the effects of preventive feedback
(produced by, e.g., radiation pressure of massive stars and local photoionization), which reduces 
the conversion efficiency of gas into stars.  

After the completion of this paper, several works have appeared in the arXiv, showing that the introduction
of preventive feedback may indeed delay the SF in simulated galaxies \citep{Kannan+2014,Sales+2014,Hopkins+2013,
Trujillo-Gomez+2013}. In particular, in the papers by \citet{Hopkins+2013} and \citet{Trujillo-Gomez+2013}, some
of their simulated galaxies, which include radiation-pressure feedback (the former authors 
include also photo-ionization and photo-electric heating), in addition to the SN feedback, 
are low-mass galaxies run to $z = 0$.  As these authors show, the inclusion of preventive feedback 
delays the SF and decouples the assembly of the galaxy from that of its DM halo 
in low-mass galaxies, helping produce dwarfs with smaller \ms-to-\mh\ ratios and higher values of 
gas fraction and sSFR at late epochs.

\section*{Acknowledgements}          
We are grateful to Dr. Peter Behroozi for providing us
his data plotted in Figures \ref{AMT} and \ref{mbar-mh}, in electronic form.    
A.G. acknowledges a PhD fellowship provided by DGEP-UNAM.
V.A. and A.G.  acknowledge CONACyT grant (Ciencia B\'asica) 167332-F, and A.V.
acknowledges PAPIIT-UNAM grant IA100212 for partial support. 

\bibliographystyle{mn2efix}

\bibliography{references}

\begin{thebibliography}{95}
\expandafter\ifx\csname natexlab\endcsname\relax\def\natexlab#1{#1}\fi

\bibitem[{{Agertz} {et~al}\mbox{.}(2013){Agertz}, {Kravtsov}, {Leitner}, \&
  {Gnedin}}]{Agertz+2013}
{Agertz} O., {Kravtsov} A.~V., {Leitner} S.~N., {Gnedin} N.~Y., 2013, \apj,
  770, 25

\bibitem[{{Agertz}, {Teyssier} \& {Moore}(2011){Agertz}, {Teyssier}, \&
  {Moore}}]{Agertz+2011}
{Agertz} O., {Teyssier} R., {Moore} B., 2011, \mnras, 410, 1391

\bibitem[{{Avila-Reese} {et~al}\mbox{.}(2011){Avila-Reese}, {Col{\'{\i}}n},
  {Gonz{\'a}lez-Samaniego}, {Valenzuela}, {Firmani}, {Vel{\'a}zquez}, \&
  {Ceverino}}]{Avila-Reese+2011b}
{Avila-Reese} V., {Col{\'{\i}}n} P., {Gonz{\'a}lez-Samaniego} A., {Valenzuela}
  O., {Firmani} C., {Vel{\'a}zquez} H., {Ceverino} D., 2011, \apj, 736, 134

\bibitem[{{Avila-Reese} {et~al}\mbox{.}(2008){Avila-Reese}, {Zavala},
  {Firmani}, \& {Hern{\'a}ndez-Toledo}}]{Avila-Reese+2008}
{Avila-Reese} V., {Zavala} J., {Firmani} C., {Hern{\'a}ndez-Toledo} H.~M.,
  2008, \aj, 136, 1340

\bibitem[{{Baldry}, {Glazebrook} \& {Driver}(2008){Baldry}, {Glazebrook}, \&
  {Driver}}]{Baldry+2008}
{Baldry} I.~K., {Glazebrook} K., {Driver} S.~P., 2008, \mnras, 388, 945

\bibitem[{{Barker} {et~al}\mbox{.}(2012){Barker}, {Ferguson}, {Irwin},
  {Arimoto}, \& {Jablonka}}]{Barker+2012}
{Barker} M.~K., {Ferguson} A.~M.~N., {Irwin} M.~J., {Arimoto} N., {Jablonka}
  P., 2012, \mnras, 419, 1489

\bibitem[{{Bauer} {et~al}\mbox{.}(2011){Bauer}, {Conselice},
  {P{\'e}rez-Gonz{\'a}lez}, {Gr{\"u}tzbauch}, {Bluck}, {Buitrago}, \&
  {Mortlock}}]{Bauer+2011}
{Bauer} A.~E., {Conselice} C.~J., {P{\'e}rez-Gonz{\'a}lez} P.~G.,
  {Gr{\"u}tzbauch} R., {Bluck} A.~F.~L., {Buitrago} F., {Mortlock} A., 2011,
  \mnras, 417, 289

\bibitem[{{Bauer} {et~al}\mbox{.}(2013){Bauer}, {Hopkins}, {Gunawardhana},
  {Taylor}, {Baldry}, {Bamford}, {Bland-Hawthorn}, {Brough}, {Brown}, {Cluver},
  {Colless}, {Conselice}, {Croom}, {Driver}, {Foster}, {Jones}, {Lara-Lopez},
  {Liske}, {L{\'o}pez-S{\'a}nchez}, {Loveday}, {Norberg}, {Owers}, {Pimbblet},
  {Robotham}, {Sansom}, \& {Sharp}}]{Bauer+2013}
{Bauer} A.~E. {et~al.}, 2013, \mnras, 434, 209

\bibitem[{{Behroozi}, {Conroy} \& {Wechsler}(2010){Behroozi}, {Conroy}, \&
  {Wechsler}}]{Behroozi+2010}
{Behroozi} P.~S., {Conroy} C., {Wechsler} R.~H., 2010, \apj, 717, 379

\bibitem[{{Behroozi}, {Wechsler} \& {Conroy}(2013){Behroozi}, {Wechsler}, \&
  {Conroy}}]{Behroozi+2013}
{Behroozi} P.~S., {Wechsler} R.~H., {Conroy} C., 2013, \apj, 770, 57

\bibitem[{{Bell} {et~al}\mbox{.}(2003){Bell}, {McIntosh}, {Katz}, \&
  {Weinberg}}]{Bell+2003b}
{Bell} E.~F., {McIntosh} D.~H., {Katz} N., {Weinberg} M.~D., 2003, \apjs, 149,
  289

\bibitem[{{Bernard} {et~al}\mbox{.}(2012){Bernard}, {Ferguson}, {Barker},
  {Irwin}, {Jablonka}, \& {Arimoto}}]{Bernard+2012}
{Bernard} E.~J., {Ferguson} A.~M.~N., {Barker} M.~K., {Irwin} M.~J., {Jablonka}
  P., {Arimoto} N., 2012, \mnras, 426, 3490

\bibitem[{{Blanton} {et~al}\mbox{.}(2005{\natexlab{a}}){Blanton}, {Lupton},
  {Schlegel}, {Strauss}, {Brinkmann}, {Fukugita}, \& {Loveday}}]{Blanton+2005a}
{Blanton} M.~R., {Lupton} R.~H., {Schlegel} D.~J., {Strauss} M.~A., {Brinkmann}
  J., {Fukugita} M., {Loveday} J., 2005{\natexlab{a}}, \apj, 631, 208

\bibitem[{{Blanton} {et~al}\mbox{.}(2005{\natexlab{b}}){Blanton}, {Schlegel},
  {Strauss}, {Brinkmann}, {Finkbeiner}, {Fukugita}, {Gunn}, {Hogg},
  {Ivezi{\'c}}, {Knapp}, {Lupton}, {Munn}, {Schneider}, {Tegmark}, \&
  {Zehavi}}]{Blanton+2005b}
{Blanton} M.~R. {et~al.}, 2005{\natexlab{b}}, \aj, 129, 2562

\bibitem[{{Bouch{\'e}} {et~al}\mbox{.}(2010){Bouch{\'e}}, {Dekel}, {Genzel},
  {Genel}, {Cresci}, {F{\"o}rster Schreiber}, {Shapiro}, {Davies}, \&
  {Tacconi}}]{Bouchet+2010}
{Bouch{\'e}} N. {et~al.}, 2010, \apj, 718, 1001

\bibitem[{{Brook} {et~al}\mbox{.}(2012){Brook}, {Stinson}, {Gibson}, {Wadsley},
  \& {Quinn}}]{Brook+2012}
{Brook} C.~B., {Stinson} G., {Gibson} B.~K., {Wadsley} J., {Quinn} T., 2012,
  \mnras, 424, 1275

\bibitem[{{Bryan} \& {Norman}(1997)}]{Bryan+1997}
{Bryan} G.~L., {Norman} M.~L., 1997, in Astronomical Society of the Pacific
  Conference Series, Vol. 123, Computational Astrophysics; 12th Kingston
  Meeting on Theoretical Astrophysics, {Clarke} D.~A., {West} M.~J., eds., p.
  363

\bibitem[{{Bryan} {et~al}\mbox{.}(2013){Bryan}, {Kay}, {Duffy}, {Schaye},
  {Dalla Vecchia}, \& {Booth}}]{Bryan+2013}
{Bryan} S.~E., {Kay} S.~T., {Duffy} A.~R., {Schaye} J., {Dalla Vecchia} C.,
  {Booth} C.~M., 2013, \mnras, 429, 3316

\bibitem[{{Ceverino} {et~al}\mbox{.}(2013){Ceverino}, {Klypin}, {Klimek},
  {Trujillo-Gomez}, {Churchill}, {Primack}, \& {Dekel}}]{Ceverino+2013}
{Ceverino} D., {Klypin} A., {Klimek} E., {Trujillo-Gomez} S., {Churchill}
  C.~W., {Primack} J., {Dekel} A., 2013, arXiv:1307.0943

\bibitem[{{Christensen} {et~al}\mbox{.}(2012){Christensen}, {Quinn},
  {Governato}, {Stilp}, {Shen}, \& {Wadsley}}]{Christensen+2012}
{Christensen} C., {Quinn} T., {Governato} F., {Stilp} A., {Shen} S., {Wadsley}
  J., 2012, \mnras, 425, 3058

\bibitem[{{Col{\'{\i}}n} {et~al}\mbox{.}(2010){Col{\'{\i}}n}, {Avila-Reese},
  {V{\'a}zquez-Semadeni}, {Valenzuela}, \& {Ceverino}}]{Colin+2010}
{Col{\'{\i}}n} P., {Avila-Reese} V., {V{\'a}zquez-Semadeni} E., {Valenzuela}
  O., {Ceverino} D., 2010, \apj, 713, 535

\bibitem[{{Conroy}, {Gunn} \& {White}(2009){Conroy}, {Gunn}, \&
  {White}}]{Conroy+2009b}
{Conroy} C., {Gunn} J.~E., {White} M., 2009, \apj, 699, 486

\bibitem[{{Conroy} \& {Wechsler}(2009)}]{Conroy+2009a}
{Conroy} C., {Wechsler} R.~H., 2009, \apj, 696, 620

\bibitem[{{Dalla Vecchia} \& {Schaye}(2012)}]{DallaVecchia+2012}
{Dalla Vecchia} C., {Schaye} J., 2012, \mnras, 426, 140

\bibitem[{{de Blok} \& {Walter}(2006)}]{deBlok+2006}
{de Blok} W.~J.~G., {Walter} F., 2006, \aj, 131, 343

\bibitem[{{De Rossi} {et~al}\mbox{.}(2013){De Rossi}, {Avila-Reese}, {Tissera},
  {Gonz{\'a}lez-Samaniego}, \& {Pedrosa}}]{deRossi+2013}
{De Rossi} M.~E., {Avila-Reese} V., {Tissera} P.~B., {Gonz{\'a}lez-Samaniego}
  A., {Pedrosa} S.~E., 2013, \mnras, 435, 2736

\bibitem[{{de Rossi}, {Tissera} \& {Pedrosa}(2010){de Rossi}, {Tissera}, \&
  {Pedrosa}}]{deRossi+2010}
{de Rossi} M.~E., {Tissera} P.~B., {Pedrosa} S.~E., 2010, \aap, 519, A89+

\bibitem[{{Duffy} {et~al}\mbox{.}(2010){Duffy}, {Schaye}, {Kay}, {Dalla
  Vecchia}, {Battye}, \& {Booth}}]{Duffy+2010}
{Duffy} A.~R., {Schaye} J., {Kay} S.~T., {Dalla Vecchia} C., {Battye} R.~A.,
  {Booth} C.~M., 2010, \mnras, 405, 2161

\bibitem[{{Dutton} {et~al}\mbox{.}(2011){Dutton}, {van den Bosch}, {Faber},
  {Simard}, {Kassin}, {Koo}, {Bundy}, {Huang}, {Weiner}, {Cooper}, {Newman},
  {Mozena}, \& {Koekemoer}}]{Dutton+2011}
{Dutton} A.~A. {et~al.}, 2011, \mnras, 410, 1660

\bibitem[{{Fakhouri}, {Ma} \& {Boylan-Kolchin}(2010){Fakhouri}, {Ma}, \&
  {Boylan-Kolchin}}]{Fakhouri+2010}
{Fakhouri} O., {Ma} C.-P., {Boylan-Kolchin} M., 2010, \mnras, 406, 2267

\bibitem[{{Ferland} {et~al}\mbox{.}(1998){Ferland}, {Korista}, {Verner},
  {Ferguson}, {Kingdon}, \& {Verner}}]{Ferland+1998}
{Ferland} G.~J., {Korista} K.~T., {Verner} D.~A., {Ferguson} J.~W., {Kingdon}
  J.~B., {Verner} E.~M., 1998, \pasp, 110, 761

\bibitem[{{Firmani} \& {Avila-Reese}(2010)}]{Firmani+2010b}
{Firmani} C., {Avila-Reese} V., 2010, \apj, 723, 755

\bibitem[{{Firmani}, {Avila-Reese} \& {Rodr{\'{\i}}guez-Puebla}(2010){Firmani},
  {Avila-Reese}, \& {Rodr{\'{\i}}guez-Puebla}}]{Firmani+2010a}
{Firmani} C., {Avila-Reese} V., {Rodr{\'{\i}}guez-Puebla} A., 2010, \mnras,
  404, 1100

\bibitem[{{Fontanot} {et~al}\mbox{.}(2009){Fontanot}, {De Lucia}, {Monaco},
  {Somerville}, \& {Santini}}]{Fontanot+2009}
{Fontanot} F., {De Lucia} G., {Monaco} P., {Somerville} R.~S., {Santini} P.,
  2009, \mnras, 397, 1776

\bibitem[{{Geha} {et~al}\mbox{.}(2006){Geha}, {Blanton}, {Masjedi}, \&
  {West}}]{Geha+2006}
{Geha} M., {Blanton} M.~R., {Masjedi} M., {West} A.~A., 2006, \apj, 653, 240

\bibitem[{{Geha} {et~al}\mbox{.}(2012){Geha}, {Blanton}, {Yan}, \&
  {Tinker}}]{Geha+2012}
{Geha} M., {Blanton} M.~R., {Yan} R., {Tinker} J.~L., 2012, \apj, 757, 85

\bibitem[{{Gilbank} {et~al}\mbox{.}(2011){Gilbank}, {Bower}, {Glazebrook},
  {Balogh}, {Baldry}, {Davies}, {Hau}, {Li}, {McCarthy}, \&
  {Sawicki}}]{Gilbank+2011}
{Gilbank} D.~G. {et~al.}, 2011, \mnras, 414, 304

\bibitem[{{Guo} {et~al}\mbox{.}(2010){Guo}, {White}, {Li}, \&
  {Boylan-Kolchin}}]{Guo+2010}
{Guo} Q., {White} S., {Li} C., {Boylan-Kolchin} M., 2010, \mnras, 404, 1111

\bibitem[{{Haardt} \& {Madau}(1996)}]{Haardt+1996}
{Haardt} F., {Madau} P., 1996, \apj, 461, 20

\bibitem[{{Hopkins} {et~al}\mbox{.}(2013){Hopkins}, {Keres}, {Onorbe},
  {Faucher-Giguere}, {Quataert}, {Murray}, \& {Bullock}}]{Hopkins+2013}
{Hopkins} P.~F., {Keres} D., {Onorbe} J., {Faucher-Giguere} C.-A., {Quataert}
  E., {Murray} N., {Bullock} J.~S., 2013, arXiv:1311.2073

\bibitem[{{Hopkins}, {Quataert} \& {Murray}(2012){Hopkins}, {Quataert}, \&
  {Murray}}]{Hopkins+2012}
{Hopkins} P.~F., {Quataert} E., {Murray} N., 2012, \mnras, 421, 3522

\bibitem[{{Hummels} \& {Bryan}(2012)}]{Hummels+2012}
{Hummels} C.~B., {Bryan} G.~L., 2012, \apj, 749, 140

\bibitem[{{Hunter}, {Elmegreen} \& {Ludka}(2010){Hunter}, {Elmegreen}, \&
  {Ludka}}]{Hunter+2010}
{Hunter} D.~A., {Elmegreen} B.~G., {Ludka} B.~C., 2010, \aj, 139, 447

\bibitem[{{Kannan} {et~al}\mbox{.}(2014){Kannan}, {Stinson}, {Macci{\`o}},
  {Hennawi}, {Woods}, {Wadsley}, {Shen}, {Robitaille}, {Cantalupo}, {Quinn}, \&
  {Christensen}}]{Kannan+2014}
{Kannan} R. {et~al.}, 2014, \mnras, 437, 2882

\bibitem[{{Karim} {et~al}\mbox{.}(2011){Karim}, {Schinnerer},
  {Mart{\'{\i}}nez-Sansigre}, {Sargent}, {van der Wel}, {Rix}, {Ilbert},
  {Smol{\v c}i{\'c}}, {Carilli}, {Pannella}, {Koekemoer}, {Bell}, \&
  {Salvato}}]{Karim+2011}
{Karim} A. {et~al.}, 2011, \apj, 730, 61

\bibitem[{{Klypin} {et~al}\mbox{.}(1999){Klypin}, {Gottl{\"o}ber}, {Kravtsov},
  \& {Khokhlov}}]{Klypin+1999}
{Klypin} A., {Gottl{\"o}ber} S., {Kravtsov} A.~V., {Khokhlov} A.~M., 1999,
  \apj, 516, 530

\bibitem[{{Klypin} \& {Holtzman}(1997)}]{Klypin+1997}
{Klypin} A., {Holtzman} J., 1997, preprint (astro-ph/9712217)

\bibitem[{{Klypin} {et~al}\mbox{.}(2001){Klypin}, {Kravtsov}, {Bullock}, \&
  {Primack}}]{Klypin+2001}
{Klypin} A., {Kravtsov} A.~V., {Bullock} J.~S., {Primack} J.~R., 2001, \apj,
  554, 903

\bibitem[{{Klypin}, {Trujillo-Gomez} \& {Primack}(2011){Klypin},
  {Trujillo-Gomez}, \& {Primack}}]{Klypin+2011}
{Klypin} A.~A., {Trujillo-Gomez} S., {Primack} J., 2011, \apj, 740, 102

\bibitem[{{Kravtsov}(2003)}]{Kravtsov2003}
{Kravtsov} A.~V., 2003, \apjl, 590, L1

\bibitem[{{Kravtsov}, {Klypin} \& {Khokhlov}(1997){Kravtsov}, {Klypin}, \&
  {Khokhlov}}]{Kravtsov+1997}
{Kravtsov} A.~V., {Klypin} A.~A., {Khokhlov} A.~M., 1997, \apjs, 111, 73

\bibitem[{{Krumholz} \& {Dekel}(2012)}]{Krumholz+2011}
{Krumholz} M.~R., {Dekel} A., 2012, \apj, 753, 16

\bibitem[{{Krumholz} \& {Tan}(2007)}]{Krumholz+2007}
{Krumholz} M.~R., {Tan} J.~C., 2007, \apj, 654, 304

\bibitem[{{Kuhlen} {et~al}\mbox{.}(2012){Kuhlen}, {Krumholz}, {Madau}, {Smith},
  \& {Wise}}]{Kuhlen+2012}
{Kuhlen} M., {Krumholz} M.~R., {Madau} P., {Smith} B.~D., {Wise} J., 2012,
  \apj, 749, 36

\bibitem[{{Lada} {et~al}\mbox{.}(2012){Lada}, {Forbrich}, {Lombardi}, \&
  {Alves}}]{Lada+2012}
{Lada} C.~J., {Forbrich} J., {Lombardi} M., {Alves} J.~F., 2012, \apj, 745, 190

\bibitem[{{Lada}, {Lombardi} \& {Alves}(2010){Lada}, {Lombardi}, \&
  {Alves}}]{Lada+2010}
{Lada} C.~J., {Lombardi} M., {Alves} J.~F., 2010, \apj, 724, 687

\bibitem[{{Lee} {et~al}\mbox{.}(2009){Lee}, {Kennicutt}, {Funes}, {Sakai}, \&
  {Akiyama}}]{Lee+2009}
{Lee} J.~C., {Kennicutt}, Jr. R.~C., {Funes} S.~J.~J.~G., {Sakai} S., {Akiyama}
  S., 2009, \apj, 692, 1305

\bibitem[{{Leitner}(2012)}]{Leitner2012}
{Leitner} S.~N., 2012, \apj, 745, 149

\bibitem[{{Liu} {et~al}\mbox{.}(2010){Liu}, {Yang}, {Mo}, {van den Bosch}, \&
  {Springel}}]{Liu+2010}
{Liu} L., {Yang} X., {Mo} H.~J., {van den Bosch} F.~C., {Springel} V., 2010,
  \apj, 712, 734

\bibitem[{{Mateo}(1998)}]{Mateo+1998}
{Mateo} M.~L., 1998, \araa, 36, 435

\bibitem[{{McGaugh}(2012)}]{McGaugh+2012}
{McGaugh} S.~S., 2012, \aj, 143, 40

\bibitem[{{McQuinn} {et~al}\mbox{.}(2010){McQuinn}, {Skillman}, {Cannon},
  {Dalcanton}, {Dolphin}, {Hidalgo-Rodr{\'{\i}}guez}, {Holtzman}, {Stark},
  {Weisz}, \& {Williams}}]{McQuinn+2010}
{McQuinn} K.~B.~W. {et~al.}, 2010, \apj, 721, 297

\bibitem[{{Moster}, {Naab} \& {White}(2013){Moster}, {Naab}, \&
  {White}}]{Moster+2013}
{Moster} B.~P., {Naab} T., {White} S.~D.~M., 2013, \mnras, 428, 3121

\bibitem[{{Moster} {et~al}\mbox{.}(2010){Moster}, {Somerville}, {Maulbetsch},
  {van den Bosch}, {Macci{\`o}}, {Naab}, \& {Oser}}]{Moster+2010}
{Moster} B.~P., {Somerville} R.~S., {Maulbetsch} C., {van den Bosch} F.~C.,
  {Macci{\`o}} A.~V., {Naab} T., {Oser} L., 2010, \apj, 710, 903

\bibitem[{{Moustakas} {et~al}\mbox{.}(2013){Moustakas}, {Coil}, {Aird},
  {Blanton}, {Cool}, {Eisenstein}, {Mendez}, {Wong}, {Zhu}, \&
  {Arnouts}}]{Moustakas+2013}
{Moustakas} J. {et~al.}, 2013, \apj, 767, 50

\bibitem[{{Munshi} {et~al}\mbox{.}(2013){Munshi}, {Governato}, {Brooks},
  {Christensen}, {Shen}, {Loebman}, {Moster}, {Quinn}, \&
  {Wadsley}}]{Munshi+2013}
{Munshi} F. {et~al.}, 2013, \apj, 766, 56

\bibitem[{{Murray}, {Quataert} \& {Thompson}(2005){Murray}, {Quataert}, \&
  {Thompson}}]{Murray+2005}
{Murray} N., {Quataert} E., {Thompson} T.~A., 2005, \apj, 618, 569

\bibitem[{{Noeske} {et~al}\mbox{.}(2007){Noeske}, {Faber}, {Weiner}, {Koo},
  {Primack}, {Dekel}, {Papovich}, {Conselice}, {Le Floc'h}, {Rieke}, {Coil},
  {Lotz}, {Somerville}, \& {Bundy}}]{Noeske+2007b}
{Noeske} K.~G. {et~al.}, 2007, \apjl, 660, L47

\bibitem[{{Pacifici} {et~al}\mbox{.}(2013){Pacifici}, {Kassin}, {Weiner},
  {Charlot}, \& {Gardner}}]{Pacifici+2013}
{Pacifici} C., {Kassin} S.~A., {Weiner} B., {Charlot} S., {Gardner} J.~P.,
  2013, \apjl, 762, L15

\bibitem[{{Papastergis} {et~al}\mbox{.}(2012){Papastergis}, {Cattaneo},
  {Huang}, {Giovanelli}, \& {Haynes}}]{Papastergis+2012}
{Papastergis} E., {Cattaneo} A., {Huang} S., {Giovanelli} R., {Haynes} M.~P.,
  2012, \apj, 759, 138

\bibitem[{{P{\'e}rez} {et~al}\mbox{.}(2013){P{\'e}rez}, {Cid Fernandes},
  {Gonz{\'a}lez Delgado}, {Garc{\'{\i}}a-Benito}, {S{\'a}nchez}, {Husemann},
  {Mast}, {Rod{\'o}n}, {Kupko}, {Backsmann}, {de Amorim}, {van de Ven},
  {Walcher}, {Wisotzki}, {Cortijo-Ferrero}, \& {collaboration6}}]{Perez+2013}
{P{\'e}rez} E. {et~al.}, 2013, \apjl, 764, L1

\bibitem[{{Rodighiero} {et~al}\mbox{.}(2010){Rodighiero}, {Cimatti},
  {Gruppioni}, {Popesso}, {Andreani}, {Altieri}, {Aussel}, {Berta},
  {Bongiovanni}, {Brisbin}, {Cava}, {Cepa}, {Daddi}, {Dominguez-Sanchez},
  {Elbaz}, {Fontana}, {F{\"o}rster Schreiber}, {Franceschini}, {Genzel},
  {Grazian}, {Lutz}, {Magdis}, {Magliocchetti}, {Magnelli}, {Maiolino},
  {Mancini}, {Nordon}, {Perez Garcia}, {Poglitsch}, {Santini},
  {Sanchez-Portal}, {Pozzi}, {Riguccini}, {Saintonge}, {Shao}, {Sturm},
  {Tacconi}, {Valtchanov}, {Wetzstein}, \& {Wieprecht}}]{Rodighiero+2010}
{Rodighiero} G. {et~al.}, 2010, \aap, 518, L25+

\bibitem[{{Rodr{\'{\i}}guez-Puebla}, {Avila-Reese} \&
  {Drory}(2013){Rodr{\'{\i}}guez-Puebla}, {Avila-Reese}, \&
  {Drory}}]{Rodriguez-Puebla+2013}
{Rodr{\'{\i}}guez-Puebla} A., {Avila-Reese} V., {Drory} N., 2013, \apj, 767, 92

\bibitem[{{Rodr{\'{\i}}guez-Puebla}
  {et~al}\mbox{.}(2011){Rodr{\'{\i}}guez-Puebla}, {Avila-Reese}, {Firmani}, \&
  {Col{\'{\i}}n}}]{Rodriguez-Puebla+2011}
{Rodr{\'{\i}}guez-Puebla} A., {Avila-Reese} V., {Firmani} C., {Col{\'{\i}}n}
  P., 2011, \rmxaa, 47, 235

\bibitem[{{Rodr{\'{\i}}guez-Puebla}, {Drory} \&
  {Avila-Reese}(2012){Rodr{\'{\i}}guez-Puebla}, {Drory}, \&
  {Avila-Reese}}]{Rodriguez-Puebla+2012}
{Rodr{\'{\i}}guez-Puebla} A., {Drory} N., {Avila-Reese} V., 2012, \apj, 756, 2

\bibitem[{{Sales} {et~al}\mbox{.}(2014){Sales}, {Marinacci}, {Springel}, \&
  {Petkova}}]{Sales+2014}
{Sales} L.~V., {Marinacci} F., {Springel} V., {Petkova} M., 2014, \mnras, in
  press, arXvi:1310.7572

\bibitem[{{Salim} {et~al}\mbox{.}(2007){Salim}, {Rich}, {Charlot},
  {Brinchmann}, {Johnson}, {Schiminovich}, {Seibert}, {Mallery}, {Heckman},
  {Forster}, {Friedman}, {Martin}, {Morrissey}, {Neff}, {Small}, {Wyder},
  {Bianchi}, {Donas}, {Lee}, {Madore}, {Milliard}, {Szalay}, {Welsh}, \&
  {Yi}}]{Salim+2007}
{Salim} S. {et~al.}, 2007, \apjs, 173, 267

\bibitem[{{Santini} {et~al}\mbox{.}(2009){Santini}, {Fontana}, {Grazian},
  {Salimbeni}, {Fiore}, {Fontanot}, {Boutsia}, {Castellano}, {Cristiani}, {de
  Santis}, {Gallozzi}, {Giallongo}, {Menci}, {Nonino}, {Paris}, {Pentericci},
  \& {Vanzella}}]{Santini+2009}
{Santini} P. {et~al.}, 2009, \aap, 504, 751

\bibitem[{{Sawala} {et~al}\mbox{.}(2013){Sawala}, {Frenk}, {Crain}, {Jenkins},
  {Schaye}, {Theuns}, \& {Zavala}}]{Sawala+2013}
{Sawala} T., {Frenk} C.~S., {Crain} R.~A., {Jenkins} A., {Schaye} J., {Theuns}
  T., {Zavala} J., 2013, \mnras, 431, 1366

\bibitem[{{Sawala} {et~al}\mbox{.}(2011){Sawala}, {Guo}, {Scannapieco},
  {Jenkins}, \& {White}}]{Sawala+2011}
{Sawala} T., {Guo} Q., {Scannapieco} C., {Jenkins} A., {White} S., 2011,
  \mnras, 413, 659

\bibitem[{{Scannapieco} {et~al}\mbox{.}(2012){Scannapieco}, {Wadepuhl},
  {Parry}, {Navarro}, {Jenkins}, {Springel}, {Teyssier}, {Carlson}, {Couchman},
  {Crain}, {Dalla Vecchia}, {Frenk}, {Kobayashi}, {Monaco}, {Murante},
  {Okamoto}, {Quinn}, {Schaye}, {Stinson}, {Theuns}, {Wadsley}, {White}, \&
  {Woods}}]{Scannapieco+2012}
{Scannapieco} C. {et~al.}, 2012, \mnras, 423, 1726

\bibitem[{{Somerville} {et~al}\mbox{.}(2008){Somerville}, {Barden}, {Rix},
  {Bell}, {Beckwith}, {Borch}, {Caldwell}, {H{\"a}u{\ss}ler}, {Heymans},
  {Jahnke}, {Jogee}, {McIntosh}, {Meisenheimer}, {Peng}, {S{\'a}nchez},
  {Wisotzki}, \& {Wolf}}]{Somerville+2008}
{Somerville} R.~S. {et~al.}, 2008, \apj, 672, 776

\bibitem[{{Stewart} {et~al}\mbox{.}(2009){Stewart}, {Bullock}, {Wechsler}, \&
  {Maller}}]{Stewart+2009}
{Stewart} K.~R., {Bullock} J.~S., {Wechsler} R.~H., {Maller} A.~H., 2009, \apj,
  702, 307

\bibitem[{{Stinson} {et~al}\mbox{.}(2006){Stinson}, {Seth}, {Katz}, {Wadsley},
  {Governato}, \& {Quinn}}]{Stinson+2006}
{Stinson} G., {Seth} A., {Katz} N., {Wadsley} J., {Governato} F., {Quinn} T.,
  2006, \mnras, 373, 1074

\bibitem[{{Stinson} {et~al}\mbox{.}(2009){Stinson}, {Dalcanton}, {Quinn},
  {Gogarten}, {Kaufmann}, \& {Wadsley}}]{Stinson+2009}
{Stinson} G.~S., {Dalcanton} J.~J., {Quinn} T., {Gogarten} S.~M., {Kaufmann}
  T., {Wadsley} J., 2009, \mnras, 395, 1455

\bibitem[{{Stinson} {et~al}\mbox{.}(2007){Stinson}, {Dalcanton}, {Quinn},
  {Kaufmann}, \& {Wadsley}}]{Stinson+2007}
{Stinson} G.~S., {Dalcanton} J.~J., {Quinn} T., {Kaufmann} T., {Wadsley} J.,
  2007, \apj, 667, 170

\bibitem[{{Teyssier}(2002)}]{Teyssier+2002}
{Teyssier} R., 2002, \aap, 385, 337

\bibitem[{{Teyssier} {et~al}\mbox{.}(2013){Teyssier}, {Pontzen}, {Dubois}, \&
  {Read}}]{Teyssier+2013}
{Teyssier} R., {Pontzen} A., {Dubois} Y., {Read} J.~I., 2013, \mnras, 429, 3068

\bibitem[{{Thompson} {et~al}\mbox{.}(2014){Thompson}, {Nagamine}, {Jaacks}, \&
  {Choi}}]{Thompson+2014}
{Thompson} R., {Nagamine} K., {Jaacks} J., {Choi} J.-H., 2014, \apj, 780, 145

\bibitem[{{Trujillo-Gomez} {et~al}\mbox{.}(2013){Trujillo-Gomez}, {Klypin},
  {Colin}, {Ceverino}, {Arraki}, \& {Primack}}]{Trujillo-Gomez+2013}
{Trujillo-Gomez} S., {Klypin} A., {Colin} P., {Ceverino} D., {Arraki} K.,
  {Primack} J., 2013, arXiv:1311.2910

\bibitem[{{Weinmann} {et~al}\mbox{.}(2012){Weinmann}, {Pasquali},
  {Oppenheimer}, {Finlator}, {Mendel}, {Crain}, \&
  {Macci{\`o}}}]{Weinmann+2012}
{Weinmann} S.~M., {Pasquali} A., {Oppenheimer} B.~D., {Finlator} K., {Mendel}
  J.~T., {Crain} R.~A., {Macci{\`o}} A.~V., 2012, \mnras, 426, 2797

\bibitem[{{Weisz} {et~al}\mbox{.}(2011){Weisz}, {Dalcanton}, {Williams},
  {Gilbert}, {Skillman}, {Seth}, {Dolphin}, {McQuinn}, {Gogarten}, {Holtzman},
  {Rosema}, {Cole}, {Karachentsev}, \& {Zaritsky}}]{Weisz+2011b}
{Weisz} D.~R. {et~al.}, 2011, \apj, 739, 5

\bibitem[{{Weisz} {et~al}\mbox{.}(2008){Weisz}, {Skillman}, {Cannon},
  {Dolphin}, {Kennicutt}, {Lee}, \& {Walter}}]{Weisz+2008}
{Weisz} D.~R., {Skillman} E.~D., {Cannon} J.~M., {Dolphin} A.~E., {Kennicutt},
  Jr. R.~C., {Lee} J., {Walter} F., 2008, \apj, 689, 160

\bibitem[{{Wise} {et~al}\mbox{.}(2012){Wise}, {Abel}, {Turk}, {Norman}, \&
  {Smith}}]{Wise+2012}
{Wise} J.~H., {Abel} T., {Turk} M.~J., {Norman} M.~L., {Smith} B.~D., 2012,
  \mnras, 427, 311

\bibitem[{{Yang} {et~al}\mbox{.}(2012){Yang}, {Mo}, {van den Bosch}, {Zhang},
  \& {Han}}]{Yang+2012}
{Yang} X., {Mo} H.~J., {van den Bosch} F.~C., {Zhang} Y., {Han} J., 2012, \apj,
  752, 41

\end{thebibliography}

\label{lastpage}

\end{document}